\documentclass[10pt,twocolumn]{article}
\usepackage{iftex}
\ifPDFTeX
  \usepackage[utf8]{inputenc}
  \usepackage[T1]{fontenc}
\fi
\usepackage{textcomp}   
\usepackage[margin=1.9cm,columnsep=0.62cm]{geometry}
\usepackage{microtype}
\usepackage{booktabs}
\usepackage{tikz}
\usetikzlibrary{shapes.geometric,positioning}
\usepackage{graphicx}
\usepackage{array}
\usepackage{tabularx}
\usepackage{longtable}
\usepackage[numbers,sort&compress]{natbib}
\usepackage[hidelinks]{hyperref}
\usepackage{cleveref}
\usepackage{stfloats}   

\usepackage{titlesec}
\titleformat{\section}{\large\bfseries}{\thesection}{0.6em}{}
\titleformat{\subsection}{\normalsize\bfseries}{\thesubsection}{0.6em}{}
\titlespacing{\section}{0pt}{0.7em}{0.3em}
\titlespacing{\subsection}{0pt}{0.55em}{0.25em}

\usepackage[font=small,labelfont=bf,skip=3pt]{caption}

\usepackage{titling}
\usepackage{colortbl}
\usepackage{tikz}
\usetikzlibrary{arrows.meta,calc,positioning,backgrounds,fit,shapes.geometric}
\usepackage{graphicx}

\definecolor{sapphire}{HTML}{0F52BA}      
\definecolor{sapphireDark}{HTML}{0B3C8C}  
\definecolor{sapphireMid}{HTML}{4A7FD4}   
\definecolor{sapphireLight}{HTML}{DCE7F8} 
\definecolor{sapphirePale}{HTML}{EEF3FC}  
\definecolor{inkGrey}{HTML}{2B2B2B}       

\newif\ifpractitioner
\IfFileExists{practitioner.flag}{\practitionertrue}{\practitionerfalse}

\titleformat{\section}{\large\bfseries\color{sapphireDark}}{\thesection}{0.6em}{}
\titleformat{\subsection}{\normalsize\bfseries\color{sapphireDark}}{\thesubsection}{0.6em}{}

\hypersetup{colorlinks=true, linkcolor=sapphireDark, citecolor=sapphireDark,
  urlcolor=sapphire, filecolor=sapphireDark}

\newcommand{\headrow}{\rowcolor{sapphire!14}}
\newcommand{\hcell}[1]{{\bfseries\color{sapphireDark}#1}}

\newcommand{\erdosmark}[1][sapphire]{%
  \begin{tikzpicture}[line cap=round, line join=round, baseline=(current bounding box.center)]
    \coordinate (TL) at (0,0.92); \coordinate (TR) at (0.66,0.92);
    \coordinate (BL) at (0,0);    \coordinate (BR) at (0.66,0);
    \draw[#1, line width=1.5pt] (TL)--(TR) (BL)--(BR) (TL)--(BR) (TR)--(BL);
    \foreach \p in {TL,TR,BL,BR}{\fill[#1] (\p) circle (2.6pt);}
  \end{tikzpicture}}

\makeatletter
\def\@titlebannertitle{\@title}
\renewcommand{\@maketitle}{%
  \vspace*{-0.6em}%
  \noindent\begin{tikzpicture}
    \node[fill=sapphirePale, draw=sapphire!35, line width=0.8pt, rounded corners=7pt,
          inner sep=13pt, text width=\dimexpr\textwidth-30pt\relax, anchor=north west] (bx) at (0,0) {%
      \begin{minipage}{\dimexpr\textwidth-34pt\relax}
        \raisebox{-0.28\height}{\resizebox{!}{0.95em}{\erdosmark[sapphire]}}\hspace{0.6em}%
        {\small\bfseries\color{sapphireDark}Erd\H{o}s Systems}\hfill
        {\small\color{sapphireMid}Automotive compliance-evidence research}\\[0.55em]
        {\LARGE\bfseries\color{sapphireDark}\@titlebannertitle\par}%
        \vspace{0.5em}
        {\normalsize\@author\par}%
        \vspace{0.15em}
        {\small\color{sapphireMid}\@date}
      \end{minipage}};
    \fill[sapphire] ($(bx.north west)+(0,-0.06)$) rectangle ($(bx.south west)+(0.11,0.06)$);
  \end{tikzpicture}\par
  \vspace{1.4em}
}
\makeatother

\newcommand{\caseinsight}[1]{%
  \ifpractitioner
    \par\vspace{2pt}\noindent
    \begin{tikzpicture}
      \node[fill=sapphirePale, draw=sapphireMid, line width=0.6pt, rounded corners=3pt,
            inner sep=5pt, text width=\dimexpr\columnwidth-14pt\relax, font=\small\itshape,
            text=inkGrey] {\textbf{\textcolor{sapphireDark}{Insight.}}~#1};
    \end{tikzpicture}\par\vspace{2pt}%
  \else
    #1%
  \fi}

\ifpractitioner
  \usepackage{fancyhdr}
\fi

\definecolor{tierAdj}{HTML}{0B3C8C}   
\definecolor{tierAck}{HTML}{4A7FD4}   
\definecolor{tierAlg}{HTML}{B7791F}   
\definecolor{tierAlgFill}{HTML}{F5E8D0} 

\newcommand{\tierswatch}[1]{\tikz{\node[fill=#1, minimum width=1.6ex, minimum height=1.6ex,
  inner sep=0pt, rounded corners=0.4pt]{};}}

\newcommand{\mode}[1]{{\bfseries\color{sapphireDark}#1}}

\newcommand{\boundary}{{\scriptsize\itshape\color{tierAlg}(boundary case)}}

\providecommand{\xcitePone}{\mbox{[companion paper, arXiv:\,XCITE-P1]}}

\usepackage{placeins}
\title{\bfseries Compliance Evidence in the Automotive Supply Chain: A Systematisation of the Quality-Document Spine and a Taxonomy of Documentation Failure Modes}

\author{Dawar Jyoti Deka \quad Nilesh Sarkar \\[0.18em]
  \texttt{dawar@erdoslab.org} \quad \texttt{nilesh@erdoslab.org}}

\date{3 July 2026}

\begin{document}
\maketitle

\begin{abstract}
\noindent The automotive industry runs on a dense, standardised chain of supplier-quality and certification evidence: production part approval packages, initial sample reports, material certificates, inspection sign-offs, and the type-approval dossiers. The chain is operationally central, yet no literature maps it as an information system, and its failures are studied as corporate misconduct rather than system outcomes. This paper systematises the chain and taxonomises its documented failures. First, it systematises the supplier-quality evidence chain across the two dominant regimes, AIAG PPAP and VDA Volume 2 PPA, organised by artefact, producer, verifier, trigger, approval state, and retention. Second, it compiles a compendium of thirteen compliance-documentation failures made public 2012 to 2024, built strictly from the public record and each classified by evidentiary status (adjudicated, company-acknowledged, or alleged), with the Japanese certification cluster of 2016 to 2024 as centrepiece. Third, it builds a failure-mode taxonomy over the compendium, dimensioned by mechanism, lifecycle locus, driver, and detection path, shown exhaustive and discriminating over the case set. Fourth, it derives evidence-architecture requirements mechanism by mechanism and sets an open-problems agenda. Across all thirteen cases, not one failure surfaced through the chain's own routine verification; the record indicts the verification layer, not only the evidence authors.

\end{abstract}

\section{Introduction}\label{sec:intro}
On 3 June 2024, Japan's Ministry of Land, Infrastructure, Transport and Tourism announced that five vehicle manufacturers, Toyota, Mazda, Yamaha Motor, Honda and Suzuki, had acknowledged irregularities in their type designation applications. The disclosures answered a ministry order of 26 January 2024 that directed 85 type-designation holders to investigate their own certification conduct. Toyota stated the same day that seven of its models had been tested using methods differing from government standards. It temporarily halted shipments, sales and Japanese production of the three still in production: the Corolla Fielder, Corolla Axio and Yaris Cross. Akio Toyoda, in remarks reported indirectly, said the company may have been too eager to complete the tests, abbreviating them as model varieties burgeoned.

Five manufacturers acknowledging irregularities on a single day is a system event, not a coincidence of misconduct. For that one day an otherwise invisible system became visible. Automotive series production is arguably the most evidence-intensive form of mass production. Behind every part and every type approval sits a standardised chain of supplier-quality and certification evidence: part approval packages, material certificates, inspection sign-offs, and the dossiers built on them (Section 4). The chain draws notice only when it breaks, and then press coverage reports a scandal, the academic literature files a governance case study, and the information system that actually failed goes unmapped.

The gap is precise. Quality-management and practitioner writing documents part approval as procedure, not as an information system with an architecture and a failure surface. The misconduct and governance literature reads the same episodes as ethics and incentive stories. Regulatory and certification studies cover what regimes require, not the document lifecycle inside firms. Business-process compliance, RegTech, and records management supply generic frameworks with no automotive evidence-chain instantiation. Section 2 surveys these five literatures and substantiates the emptiness of their intersection.

This paper fills the gap with four contributions. C1: a systematisation of the automotive supplier-quality evidence chain across AIAG PPAP and VDA Volume 2 PPA (the PPF procedure in German usage), organised by artefact, producer, verifier, trigger, approval state, and retention obligation, within APQP and IATF 16949 (Section 4, Tables 1 and 2). C2: a verified compendium of thirteen compliance-documentation failures made public 2012-2024, classified under a three-way evidentiary discipline (adjudicated, company-acknowledged, alleged) and built strictly from the public record (Section 5, Table 3, Appendix B). C3: a failure-mode taxonomy over the compendium, dimensioned by mechanism, lifecycle locus, driver, and detection path, demonstrated exhaustive and discriminating over the case set (Section 6, \Cref{fig:taxonomy} and \Cref{fig:matrix}). C4: evidence-architecture implications mapping each failure mechanism to the property that would counter it, and an open-problems agenda (Sections 8 and 9).

A companion manuscript covers the EV-acute regulatory layer in India; this paper imports its documentation-burden lens once, in Section 7 \xcitePone.

Section 2 fixes scope and engages the five literatures. Section 3 summarises the assurance regimes that make the evidence mandatory. Section 4 systematises the chain. Section 5 presents the failure record and Section 6 the taxonomy over it. Section 7 reads the record through the burden lens, Section 8 derives evidence-architecture requirements, Section 9 states open problems, and Section 10 concludes.

\section{Background, scope, and related work}\label{sec:background}
\textbf{Scope.} This paper studies the evidence chain of automotive series production: the artefacts by which a supplier shows an OEM that a part and its process merit approval, and the OEM shows a regulator that a vehicle type conforms. It is global and regime-aware, covering type-approval and self-certification jurisdictions (Section 3), passenger and commercial vehicles. Development prototypes, aftermarket parts and recall logistics fall outside scope except where a case touches them. The unit of analysis is the document: its producer, verifier, trigger, approval state, and retention.

\textbf{Related work.} Five literatures border this object; each stops at a different wall. The quality-management and practitioner literature knows the artefacts and treats them as procedure. Doshi and Desai analyse PPAP in four SMEs as a route to quality improvement \cite{DoshiDesai2016}. Lafayette, Li and Webster risk-score by logistic regression when PPAP is required \cite{Lafayette2017}. Se\v{n}ov\'{a} et al. place PPAP inside the ISO/TS 16949 process approach, the series-parts approval mechanism \cite{Senova2021}. PPAP and APQP appear as processes to implement, algorithms to optimise or risks to score, never as an information system with producers, verifiers, approval states and a failure surface. The corporate-misconduct and governance literature knows the failures but the artefacts least of all. Hara and Fujimura attribute quality-data falsification at Japanese firms to inadequate technological capability and organisational myopia \cite{HaraFujimura2024}. Su et al. model Japanese misconduct as competing governance logics \cite{Su2025}. On Dieselgate, Rhodes reads a crisis of corporate sovereignty \cite{Rhodes2016}, Mujkic and Klingner a failure of leadership and accountability \cite{MujkicKlingner2018}, and Li et al. industry-wide corporate fraud \cite{Li2018}; Greve's reference framework treats misconduct generically \cite{Greve2010}. None analyses the compliance-evidence chain as an information system or taxonomises documentation failure modes. Regulatory and certification studies stop at the firm's boundary. Mashaw and Harfst trace US auto-safety regulation under self-certification, NHTSA's shift from rulemaking toward recall-based enforcement \cite{MashawHarfst1990}. Vodi\v{c}ka asks whether EU type approval under Regulation (EU) 2018/858 can prevent another Dieselgate \cite{Vodicka2024}. Haeckel et al. automate one conformity-of-production verification step at BMW Group \cite{Haeckel2025}. Conformity of production appears as a legal obligation, not an information flow; none systematises the evidence chain or taxonomises its failures. The information-systems literature has the method without the object. Business-process compliance, surveyed by Hashmi et al., formalises business-process checking against regulatory obligations, domain-generically \cite{Hashmi2018}. RegTech surveys are finance-centred: a 2025 review of 59 studies frames compliance, supervision and reporting in finance \cite{Bagherifam2025}. Document AI is surveyed over models, tasks and benchmarks with no domain system in view \cite{Cui2021}. Records management supplies the general theory of records as evidence. ISO 15489-1:2016 defines principles for creating, capturing and managing records in any format \cite{ISO15489-1-2016}. InterPARES targets long-term preservation of authentic digital records, domain-generically \cite{InterPARES}.

The first objection is that the scandals are already studied. They are, but as misconduct, not as evidence-system failures: Hara and Fujimura through capability and myopia \cite{HaraFujimura2024}, Su et al. through governance logics \cite{Su2025}, Rhodes as corporate sovereignty under threat \cite{Rhodes2016}. Siano et al. taxonomise its communication as greenwashing; the one taxonomy here classifies communication forms, not documentation-failure mechanisms \cite{Siano2017}. Which artefact was falsified, at which lifecycle point, under whose signature, detected how: that is never the object. This paper studies the system, not the sin.

The second objection is that this is records management or business-process compliance under another name. Those fields supply concepts this paper reuses, but are domain-generic where this paper is domain-specific. Business-process-compliance checking detects procedural deviation; a fabricated value inside a procedurally complete record is conformant by construction, a limit Sections 5 and 6 make concrete. ISO 15489 prescribes trustworthy-record principles for any domain and maps no industrial evidence chain \cite{ISO15489-1-2016}. InterPARES preserves authentic digital records without a certification instantiation \cite{InterPARES}. Intelligent-document-processing reviews cover generic forms and invoices with no compliance-evidence chain \cite{Cutting2021}. This paper supplies the missing instantiation and taxonomy.

The gap is survey-scoped, not a universal negative. Across the five literatures surveyed (53 works spanning quality management, misconduct, regulatory and certification studies, information systems, and records management), none both systematises the automotive supplier-quality evidence chain as an information system and offers a failure-mode taxonomy over it. Targeted searches under two distinct strategies found no academic systematisation of the Japanese certification-inspection cases as of the survey date. Appendix D tabulates the eight nearest neighbours and states each work's delta.

\textbf{Division of territory.} A companion manuscript, Paper 1 of this pair, holds the adjacent territory: the compliance-document lifecycle model, the India-focused EV-acute regulatory stack, and the documentation-exergy-destruction lens, which this paper does not re-derive or claim \xcitePone. The evidence-chain systematisation and the failure-mode taxonomy are this paper's territory; Section 6 borrows the lifecycle-locus dimension and Section 7 imports the lens, each with a single citation.

\section{Assurance regimes in brief}\label{sec:regimes}
Regulatory and case record stated as of 3 July 2026.

The evidence chain of Section 4 exists because market access is conditioned on proof of conformity, which the two dominant regime families demand at different points in a vehicle's life.

Type-approval regimes verify before market entry. The UNECE 1958 Agreement establishes harmonised UN Regulations with reciprocal recognition: an item type-approved by one contracting party is held to conform to the legislation of all parties applying it. EU whole-vehicle type approval operates under Regulation (EU) 2018/858, which lays down administrative and technical requirements for approving and placing on the market all new vehicles, systems, components and separate technical units. Japan grants type designation for mass-produced vehicles under Article 75 of the Road Transport Vehicle Act, administered by MLIT. Designation rests on review of sample vehicles and documents plus an audit of the manufacturer's quality control system; the manufacturer must then inspect each produced vehicle and issue a completion inspection certificate that substitutes for presenting the vehicle to the state. India implements type approval and conformity of production under the Central Motor Vehicles Rules, with testing agencies specified under Rule 126, ARAI among them, certifying compliance.

Self-certification inverts the sequence. A US manufacturer must itself certify that each vehicle complies with the Federal Motor Vehicle Safety Standards, shown by an affixed label, with no pre-market government approval; NHTSA enforces post hoc through compliance testing and defect investigation. The TREAD Act of 2000 added an early-warning layer: manufacturers must report to NHTSA possible safety-related defects and non-compliances, including foreign safety recalls. US self-certification governs safety only. For emissions, Clean Air Act sections 203 and 206 bar sale of a new vehicle unless an EPA-issued certificate of conformity covers it, and EPA conducts confirmatory testing of manufacturer-submitted emissions data: a pre-market approval, not self-certification.

Approval does not end the obligation. Under the 1958 Agreement an authority must be satisfied, before granting approval, with arrangements for ensuring conformity with the approved type. Under Regulation (EU) 2018/858 Article 31 the granting authority must verify continuing conformity of production, including checks or tests on samples and verification of a statistically relevant number of certificates of conformity. Conformity of production is thus a standing obligation, not a one-time dossier.

The regime shapes the failure surface. Type approval concentrates evidentiary risk pre-market, in the procedural record of prescribed tests, delegated inspections and approval dossiers; self-certification concentrates it post-market, in the reporting stream the regulator mines. The record of Section 5 breaks along this line; its US cases (Volkswagen, Hyundai/Kia) are emissions cases, under the EPA certificate-of-conformity regime rather than FMVSS self-certification.

After the 2024 cases, a December 2024 expert panel report led MLIT to amend the type designation rules, adding post-designation sampling tests of production vehicles from 1 April 2026, all stages now in force. A currency sweep found no further Japanese certification disclosures. In the EU, Euro 7 requires refusal of non-compliant new M1 and N1 emission type approvals from 29 November 2026; NHTSA's 2026 notices modernising FMVSS 102 and 135 for automated driving systems remain proposals.

\section{The supplier-quality evidence chain, systematised}\label{sec:chain}
\begin{figure*}[t]\centering
\resizebox{\textwidth}{!}{
\begin{tikzpicture}[
  font=\sffamily\small,
  fam/.style={draw=sapphire, line width=0.9pt, rounded corners=3pt, fill=sapphirePale,
              minimum width=3.15cm, minimum height=1.75cm, align=center, text=inkGrey,
              inner sep=4pt},
  gate/.style={draw=sapphireDark, line width=1.1pt, rounded corners=3pt, fill=sapphireLight,
               minimum width=3.0cm, minimum height=1.75cm, align=center, text=sapphireDark},
  flow/.style={-{Stealth[length=3mm,width=2.2mm]}, line width=1pt, sapphireDark},
  vda/.style={draw=sapphireMid, line width=0.8pt, dashed, rounded corners=3pt,
              fill=white, align=center, text=inkGrey, font=\sffamily\scriptsize, inner sep=4pt},
  lgnd/.style={font=\sffamily\scriptsize, text=inkGrey}
]

\node[font=\sffamily\small\bfseries, text=sapphireDark] (hdr) at (6.9,2.05)
  {PPAP submission: up to 18 elements, five submission levels (Level 3 the common default)};

\node[fam] (g1) at (0,0)
  {\textbf{Design \& change}\\ \textbf{authority}\\[2pt]
   \scriptsize design records,\\ \scriptsize engineering-change docs,\\
   \scriptsize customer approval};
\node[fam, right=0.5cm of g1] (g2)
  {\textbf{Risk \& process}\\ \textbf{planning}\\[2pt]
   \scriptsize DFMEA, process flow,\\ \scriptsize PFMEA, control plan};
\node[fam, right=0.5cm of g2] (g3)
  {\textbf{Measurement \&}\\ \textbf{capability}\\[2pt]
   \scriptsize MSA, dimensional,\\ \scriptsize material/performance,\\
   \scriptsize SPC/Cpk, lab docs};
\node[fam, right=0.5cm of g3] (g4)
  {\textbf{Appearance \&}\\ \textbf{samples}\\[2pt]
   \scriptsize AAR, sample/master\\ \scriptsize parts, checking aids};
\node[gate, right=0.6cm of g4] (psw)
  {\textbf{PSW}\\[2pt] \scriptsize part submission\\ \scriptsize warrant\\[2pt]
   \scriptsize \textbf{disposition}\\ \scriptsize \textbf{gates shipment}};

\draw[flow] (g1) -- (g2);
\draw[flow] (g2) -- (g3);
\draw[flow] (g3) -- (g4);
\draw[flow] (g4) -- (psw);

\node[lgnd, right=0.5cm of psw, align=left] (disp)
  {\textcolor{sapphireDark}{$\rightarrow$ approved}\\
   \textcolor{sapphireMid}{$\rightarrow$ interim}\\
   \textcolor{tierAlg}{$\rightarrow$ rejected}};
\draw[flow] (psw) -- (disp);

\node[vda, below=1.05cm of g1, minimum width=3.15cm] (v1)
  {VDA 2 (PPF):\\ \textit{Vorlagestufen} 1--3};
\node[vda, below=1.05cm of g2, minimum width=3.15cm] (v2)
  {two-stage risk \&\\ process analysis};
\node[vda, below=1.05cm of g3, minimum width=3.15cm] (v3)
  {capability \&\\ measurement proof};
\node[vda, below=1.05cm of g4, minimum width=3.15cm] (v4)
  {sample \& appearance\\ release};
\node[vda, below=1.05cm of psw, minimum width=3.0cm] (v5)
  {PPF cover sheet:\\ release decision};

\foreach \a/\b in {g1/v1, g2/v2, g3/v3, g4/v4, psw/v5}{
  \draw[sapphireMid, line width=0.7pt, dashed, {Stealth[length=2mm]}-{Stealth[length=2mm]}]
    (\a.south) -- (\b.north);}

\node[font=\sffamily\small\bfseries, text=sapphireMid, rotate=90]
  at ($(v1.west)+(-0.4,0)$) {VDA-PPF (Europe)};
\node[font=\sffamily\small\bfseries, text=sapphire, rotate=90]
  at ($(g1.west)+(-0.4,0)$) {PPAP (AIAG)};

\node[lgnd, anchor=north west] at ($(v1.south west)+(0,-0.15)$)
  {Dashed links mark clause-level equivalence between the AIAG and VDA release regimes; the two are aligned, not identical.};

\end{tikzpicture}}
\caption{The inherited quality-document spine. PPAP's up-to-eighteen elements group into four evidence families that flow into the PSW disposition gating shipment; the VDA-PPF (VDA~2) lane shows the aligned European release regime. The two regimes are equivalent at clause level, not identical.}\label{fig:spine}
\end{figure*}
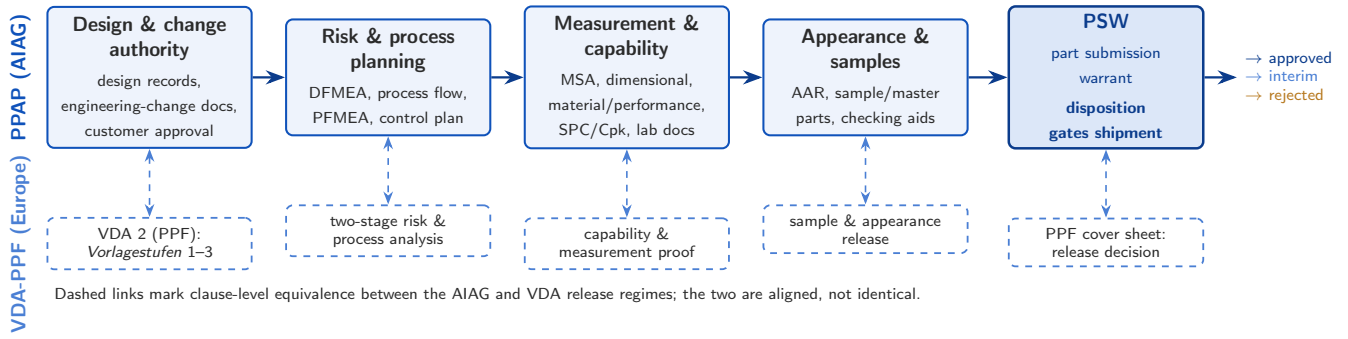
\begin{table*}[t]
\centering
\caption{The supplier-quality evidence chain by artefact family. Governing references are normative standards or instruments; verifier is the party whose sign-off gives the artefact its evidentiary force.}
\label{tab:chain}
\footnotesize
\begin{tabular}{@{}p{2.9cm}p{3.1cm}p{4.6cm}p{2.9cm}p{2.6cm}@{}}
\toprule
\headrow \hcell{Artefact} & \hcell{Governing reference} & \hcell{Purpose} & \hcell{Trigger} & \hcell{Verifier} \\
\midrule
PPAP package with Part Submission Warrant (PSW) & AIAG PPAP, 4th edn & Demonstrate that the production process meets design-record and specification requirements & New part; design, process, supplier, or site change; tooling inactive 12 months & Customer quality; disposition recorded on the PSW \\
PPA/PPF submission (traditionally ISIR/EMPB) & VDA Volume 2, 7th edn (2026) & Production process and product approval towards German customers & Scope, content, and schedule per the PPA agreement & Customer per PPA agreement \\
Control plan & AIAG Control Plan (CP-1) & Document planned process controls per part and process step; living document & Created at launch; updated with process changes & Customer review at PPAP; auditors \\
Design and process FMEA & AIAG \& VDA FMEA Handbook (2019) & Structured risk analysis of failure modes feeding the control plan & New or changed design or process & Cross-functional review; customer at submission \\
MSA study (Gage R\&R) & AIAG MSA & Evidence that measurement systems used for acceptance are adequate & Gauges named in the control plan, at PPAP & Supplier quality; customer at submission \\
Certificate of conformity (CoC) & EN 10204 type 2.1; FAR 52.246-15 as the procurement pattern & Supplier's declaration that the delivery conforms to the order & Each delivery or order & None beyond the declaring supplier \\
Material inspection certificate & EN 10204 types 3.1 and 3.2 & Lot-specific test results for metallic products & Each heat or production lot & 3.1: manufacturer's inspector, independent of production; 3.2: a party independent of the manufacturer (purchaser's representative or third-party body) countersigns \\
NCR and 8D report & IATF 16949 cl.\ 8.7, 10.2; VDA 8D volume & Document nonconformity, containment, root cause, and corrective action & Nonconforming product; customer complaint & Customer quality for customer-facing 8Ds \\
Type-approval and conformity-of-production evidence & UNECE 1958 Agreement; Reg.\ (EU) 2018/858; MLIT type designation; FMVSS self-certification & Regulator-facing evidence that the product and its ongoing production conform & Type approval; ongoing production; regulator request & Approval authority or technical service; in the US, post-hoc enforcement \\
\bottomrule
\end{tabular}
\end{table*}

\begin{table*}[t]
\centering
\caption{The two dominant production-approval regimes compared. AIAG PPAP retains fixed submission levels and an interim state; VDA Volume 2 eliminated both in its 6th edition (2020) in favour of a negotiated PPA agreement, and its 7th edition (2026) scopes the procedure by product risk.}
\label{tab:aiagvda}
\footnotesize
\begin{tabular}{@{}p{2.6cm}p{6.6cm}p{6.4cm}@{}}
\toprule
\headrow \hcell{} & \hcell{AIAG PPAP (4th edn)} & \hcell{VDA Volume 2 PPA (7th edn, 2026)} \\
\midrule
Evidence package & 18 defined elements, from design records, PFMEA, and control plan through MSA studies, initial process studies, and the PSW & PPA deliverables agreed per part; traditionally the initial sample inspection report (ISIR/EMPB), still current usage in German industry \\
Gating artefact & Part Submission Warrant: supplier-signed summary carrying reason for submission and the customer's disposition & PPA release against the PPA agreement; the ISIR/EMPB is the traditional evidence instrument \\
Producer & Supplier, as output of APQP phase 4 (product and process validation) & Supplier (the organisation), per the PPA procedure from tier n to tier n\textminus 1 \\
Verifier of record & The customer's quality organisation, whose disposition on the PSW is the approval act & The customer, per the PPA agreement \\
Submission depth & Levels 1-5 at customer discretion; Level 3 (samples plus complete supporting data) the common default; OEM overlays exist (Ford Phased PPAP; GM mandates the AIAG manual) & Submission levels eliminated in the 6th edition (2020); scope, content, and schedule agreed between organisation and customer in the PPA agreement \\
Approval states & Approved; interim approval (a manual-defined disposition permitting shipment for a limited time or quantity; whether it is granted is customer-discretionary: some customers grant none, Ford requires an approved WERS alert as the shipping authorisation); rejected & Conditional approvals eliminated in the 6th edition (2020); release per the PPA agreement \\
Triggers & New part; product design change; process change (method, tooling, location, inspection criteria); supplier or material-source change; tooling inactive 12 months & Coordinated with the customer under the PPA agreement; the 7th edition scopes the procedure by product risk level \\
Capability convention & Initial process studies against the customary 1.67 threshold (Ford CSR: Ppk $>$ 1.67 for initial and final capability) & Per PPA agreement \\
Typical retention & Production-active period plus service plus one calendar year unless the customer or a regulator specifies longer (IATF 16949 cl.\ 7.5.3.2.1; GM delegates periods to GMW15920) & Same IATF 16949 record-control umbrella; periods per customer agreement \\
Revalidation & Customer-specific: GM requires an annual layout inspection of all dimensions on at least 5 parts; Ford requires the annual layout in the control plan & Per PPA agreement \\
Convergence & \multicolumn{2}{p{13.2cm}}{Joint AIAG \& VDA FMEA Handbook (June 2019) replaced the separate methodologies; the VDA 2 7th edition foreword records that harmonisation with AIAG PPAP was discussed, with a common international standard as a long-term aim.} \\
\bottomrule
\end{tabular}
\end{table*}

This section delivers contribution C1: the supplier-to-OEM evidence chain, organised by artefact, producer, verifier, trigger, approval state, and retention obligation, and summarised by artefact family in Table 1. The chain operates under IATF 16949:2016, the automotive quality-management-system standard supplementing ISO 9001:2015; with applicable customer-specific requirements (CSRs) it defines the system within which production part approval operates.

The Production Part Approval Process (PPAP) sits within the APQP framework, in phase 4, product and process validation. It is the evidence package by which the supplier demonstrates that engineering design-record and specification requirements are met by its manufacturing process. The AIAG package comprises 18 elements, tabulated with their producers and verifiers in Appendix A. \Cref{fig:spine} traces this spine end to end: the element families converge on the PSW disposition that gates production, with the VDA PPA procedure running as an equivalence lane alongside.

PPAP defines five submission levels governing how much of the package the customer sees; the level is at the customer's discretion, with Level 3 the common default, and Table 2 lists them. The Part Submission Warrant (PSW) is the gating sign-off, required at every level. It carries the reason for submission, the authorised supplier signature, and the customer's disposition field, recording approval or rejection prior to production.

Three dispositions exist: approved, interim approval, and rejected. Interim approval permits shipment for a limited time or piece quantity while full approval is pending; rejected submissions require correction and resubmission. Granting interim approval is itself discretionary: Ford's CSR requires an approved WERS alert, the alert being the authorisation to ship. Table 2 lists the submission triggers: a new part; a design, process, supplier, or material-source change; or long-inactive tooling.

Initial process studies are judged against the customary capability threshold of 1.67; Ford's CSR states it as Ppk above 1.67. Industry sources report common rejection causes: unsigned or incomplete warrants, mismatched drawing revisions, incomplete dimensional coverage, and dimensional data without supporting capability studies. Approval does not end the obligation: annual layout inspection and revalidation are imposed as CSRs, not by the base manual. Retention runs under IATF 16949 record control plus CSRs, keeping approval records while the part is active for production and service plus one calendar year, unless a customer or regulatory agency specifies longer.

The German counterpart is VDA Volume 2 PPA (the PPF procedure in German usage), ``Quality Assurance of Supplies'', the functional counterpart of AIAG PPAP, current in its 7th revised edition of January 2026. Its traditional evidence instrument is the initial sample inspection report (ISIR; German EMPB), though current editions frame the evidence as PPA deliverables. The structural contrast is sharp: the 6th edition (2020) eliminated submission levels and conditional approvals in favour of a negotiated PPA agreement, and the 7th edition (2026) scopes the procedure by product risk level. Convergence is under way at the artefact level: AIAG and VDA jointly published the harmonised FMEA Handbook in June 2019, replacing separate methodologies with a common seven-step approach. Table 2 and Appendix C set the two regimes side by side.

The approval package travels with a supporting cast (Table 1). Control plans document the planned controls, measurements, and monitored parameters per process step, a living document under AIAG's Control Plan manual (CP-1, 2024). MSA underwrites the numbers: IATF 16949 requires Gage R\&R-type studies of each measurement system in the control plan, commonly required in PPAP submissions. Calibration and verification records are mandatory quality records, down to as-found out-of-specification readings and customer notification where suspect product shipped.

Between firms, conformity travels as certificates. A certificate of conformity (CoC) is the supplier's declaration that delivered goods conform to the order; US acquisition practice formalises the pattern as FAR clause 52.246-15. This supplier CoC, a declaration accompanying deliveries, is distinct from the vehicle certificate of conformity issued under EU whole-vehicle type approval, which Section 3 covers. For metallic products, EN 10204 grades the strength of such evidence: types 2.1 and 2.2 are manufacturer's declarations, while inspection certificates 3.1 and 3.2 rest on specific inspection of the delivered lot. A type 3.1 certificate is validated by the manufacturer's authorised inspection representative, who must be independent of the manufacturing department. A type 3.2 certificate additionally requires countersignature by a party independent of the manufacturer, the purchaser's representative or a third-party body that witnesses the tests. The graduated independence is the design principle: 3.2 exists so that a purchaser need not take the producer's word for its own metallurgy. Section 5 shows that principle tested at scale.

When product or evidence fails, the failure itself becomes a controlled record. A nonconformance report (NCR) documents and resolves deviations from requirements; IATF 16949, with ISO 9001, requires nonconformities, actions taken, and corrective-action results to be documented. The dominant report format is 8D, which Ford developed in the 1980s and the VDA standardises in a dedicated volume (2018).

Beyond the OEM, the chain continues towards the regulator as conformity of production (CoP). Under the 1958 Agreement, an approval authority must be satisfied that arrangements exist to ensure product conformity with the approved type before approval is granted. Under Regulation (EU) 2018/858 Article 31, the granting authority must verify continued conformity with the approved type, including sample checks and, for whole-vehicle approvals, verification of certificates of conformity. The supplier-quality spine and the regulator-facing dossier are one chain: the same production evidence serves two verifiers, the customer and the state.

The chain's checking stages deserve a collective name. We define the verification layer as the chain's routine, document-facing checking stages: customer disposition of submission packages, receiving-side certificate acceptance, third-party certification audit of the management system, and the manufacturer's delegated final-inspection sign-off. Regulator enforcement, meaning on-site inspections and confirmatory or audit testing, and ad-hoc engineering measurement by customers sit outside the layer: they are the enforcement backstop, not routine verification. Table 1's final column lists the designated verifier per artefact. Sections 6 and 7 examine how often this layer, rather than self-report or incident, actually detects failure.

Last, the systems reality beneath the paper chain. Analyst research reports quality-critical data fragmented across disconnected systems: manufacturing-operations (MES/MOM) and quality-management systems are often poorly connected, and low-maturity organisations hold much critical data in spreadsheets, emails, and local applications. In LNS Research's quality-management survey, integration of quality systems with disparate data sources was the single most-cited challenge in meeting quality objectives, cited by 38 per cent of respondents. A May 2021 InfinityQS poll of 215 tier-one and tier-two automotive suppliers found 55 per cent still conducting capability studies in Excel to satisfy IATF 16949 SPC requirements. From this fragmentation we infer, marking the inference as our own rather than a verified finding, that approvals routinely decouple from the artefacts they authorise: the disposition lives in one system or inbox while the evidence lives in another. The cost of that decoupling surfaces throughout the failure record that follows.

\section{The failure record}\label{sec:failures}
\begin{figure*}[t]\centering
\resizebox{\textwidth}{!}{
\begin{tikzpicture}[
  font=\sffamily\scriptsize,
  ev/.style={circle, inner sep=1.7pt},
  stem/.style={line width=0.6pt, gray!55},
  lab/.style={align=center, text=inkGrey, inner sep=1.5pt, text width=2.0cm},
  lane/.style={font=\sffamily\scriptsize\bfseries}
]

\draw[line width=1pt, inkGrey, -{Stealth[length=2.6mm]}] (-0.5,0) -- (15.0,0);
\foreach \yr in {2012,2013,2014,2015,2016,2017,2018,2019,2020,2021,2022,2023,2024}{
  \pgfmathsetmacro\xx{(\yr-2012)*1.18}
  \draw[inkGrey, line width=0.7pt] (\xx,0.09) -- (\xx,-0.09);
  \node[text=inkGrey, font=\sffamily\scriptsize\bfseries, below=5pt] at (\xx,0) {\yr};}

\node[lane, text=sapphireDark, anchor=west] at (-0.5,3.7) {Japan certification cluster};
\node[lane, text=inkGrey,      anchor=west] at (-0.5,-2.7) {global comparators};

\node[ev, fill=tierAck] (b) at (4.72,0) {};
\draw[stem] (b) -- ++(0,0.8) node[lab, above] {2016\\ Mitsubishi\\ fuel-economy (M1)};
\node[ev, fill=tierAck] (c) at (5.35,0) {};
\draw[stem] (c) -- ++(0,2.7) node[lab, above] {2016\\ Suzuki test\\ method (M2)};
\node[ev, fill=tierAdj] (d) at (5.9,0) {};
\draw[stem] (d) -- ++(0,1.75) node[lab, above] {2017\\ Kobe Steel\\ (supplier, M4)};
\node[ev, fill=tierAck] (e) at (6.5,0) {};
\draw[stem] (e) -- ++(0,3.35) node[lab, above] {2017--18\\ Nissan, Subaru\\ uncertified\\ inspectors (M3)};
\node[ev, fill=tierAdj] (f) at (11.8,0) {};
\draw[stem] (f) -- ++(0,0.8) node[lab, above] {2022\\ Hino engine\\ data (M1)};
\node[ev, fill=tierAck] (g) at (12.98,0) {};
\draw[stem] (g) -- ++(0,2.7) node[lab, above] {Dec 2023\\ Daihatsu 174\\ irregularities (M1)};
\node[ev, fill=tierAck] (h) at (13.75,0) {};
\draw[stem] (h) -- ++(0,1.75) node[lab, above] {Jan 2024\\ Toyota Ind.\\ engines (M2)};
\node[ev, fill=tierAck] (i) at (14.15,0) {};
\draw[stem] (i) -- ++(0,3.35) node[lab, above, xshift=0.25cm]
  {Jun 2024\\ five makers\\ named (M2)};

\node[ev, fill=tierAdj] (r) at (0.15,0) {};
\draw[stem] (r) -- ++(0,-2.6) node[lab, below] {2012--14\\ Hyundai/Kia\\ fuel-economy (M2)};
\node[ev, fill=tierAck] (p) at (1.35,0) {};
\draw[stem] (p) -- ++(0,-1.05) node[lab, below] {2013\\ GM India\\ Tavera (M2)};
\node[ev, fill=tierAdj] (q) at (3.65,0) {};
\draw[stem] (q) -- ++(0,-1.7) node[lab, below]
  {2015\\ \textbf{VW} defeat\\ device\\ ({\color{tierAlg}M5 boundary})};
\node[ev, fill=tierAdj] (a) at (5.9,0) {};
\draw[stem] (a) -- ++(0,-2.6) node[lab, below] {2016--17\\ Takata airbag\\ (supplier, M1)};

\node[ev, fill=tierAdj] (L1) at (9.6,-1.7) {};
\node[font=\sffamily\scriptsize, text=inkGrey, anchor=west] at (9.75,-1.7) {adjudicated};
\node[ev, fill=tierAck] (L2) at (11.6,-1.7) {};
\node[font=\sffamily\scriptsize, text=inkGrey, anchor=west] at (11.75,-1.7) {acknowledged};
\node[ev, fill=tierAlg] (L3) at (13.7,-1.7) {};
\node[font=\sffamily\scriptsize, text=inkGrey, anchor=west] at (13.85,-1.7) {alleged};

\end{tikzpicture}}
\caption{The Japanese certification cluster, 2016 to 2024 (above the axis), with global comparators (below). Marker colour keys the R5 evidence tier; mode labels key to Figure~\ref{fig:taxonomy}.}\label{fig:japantimeline}
\end{figure*}
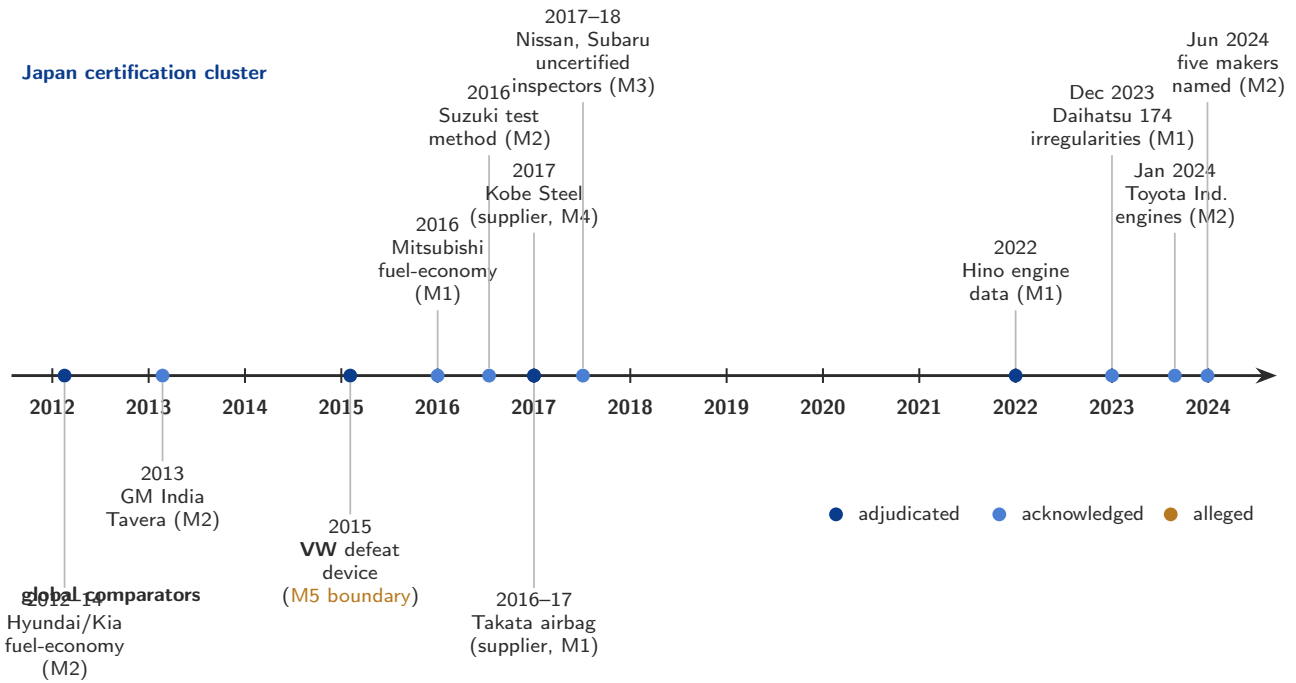
\begin{table*}[tp]
\centering
\caption{The failure compendium, ordered by the date each documentation failure became public (2012-2024); the year span runs from first public disclosure to final resolution in the record, and underlying conduct reaches back further in several cases, as the case descriptions state. Category per the evidentiary classification (adjudicated: instrument exists; acknowledged: company admission; nothing here is merely alleged). Mechanisms M1-M5 per Section~\ref{sec:taxonomy}.}
\label{tab:compendium}
\scriptsize
\begin{tabular}{@{}p{1.55cm}p{0.85cm}p{1.0cm}p{1.35cm}p{4.1cm}p{0.35cm}p{1.85cm}p{4.35cm}@{}}
\toprule
\headrow \hcell{Case} & \hcell{Years} & \hcell{Country} & \hcell{Tier} & \hcell{What evidence broke} & \hcell{M} & \hcell{How discovered} & \hcell{Consequence} \\
\midrule
Hyundai/ Kia & 2012- 2014 & KR/US & \tierswatch{tierAdj}~Adjudicated &Fuel-economy figures certified to EPA overstated economy relative to EPA's own test results & M2 & Regulator inspection & Ratings lowered and cars relabelled (Nov 2012); 2014 consent decree: USD 100M civil penalty, forfeiture of 4.75M GHG credits, compliance measures \\
GM India (Tavera) & 2013 & India & \tierswatch{tierAck}~Acknowledged &Prepared, better-performing units presented at conformity-of-production emissions verification; measurements real, specimens unrepresentative & M2 & Whistleblower or internal audit escalation & Recall of 114,000 (GM figure; ministerial 160,000 disputed); production halt; dismissals; inter-ministerial panel reportedly recommended procedural change \\
Volkswagen & 2015- 2017 & US & \tierswatch{tierAdj}~Adjudicated &No document falsified: defeat-device software met NOx limits only under test conditions, decoupling the tested from the certified artefact & M5 & Third-party or customer discovery & Three-felony plea; USD 2.8B criminal penalty, USD 1.5B civil; up to USD 14.7B June 2016 settlements; ${\sim}$11M vehicles worldwide per VW \\
Mitsubishi & 2016- 2017 & Japan & \tierswatch{tierAck}~Acknowledged &Running-resistance values in certification testing presented fuel economy better than actual; non-prescribed coasting method since 1991 beneath & M1 & Third-party or customer discovery & Production and sales halt on affected minicars; CAA surcharge order of 27 Jan 2017 (485.07M yen; follow-on orders June 2017); MLIT industry-wide instruction of 18 May 2016 \\
Suzuki & 2016 & Japan & \tierswatch{tierAck}~Acknowledged &Driving-resistance data real but produced by a non-prescribed component-accumulation route instead of the prescribed coasting test; values not overstated & M2 & Self-report & Report to MLIT of 31 May 2016 naming 26 models; re-test programme indicated catalogue figures not overstated; no administrative or judicial action in the record \\
Takata & 2016- 2017 & Japan/ US & \tierswatch{tierAdj}~Adjudicated &False inflator test reports supplied to automaker customers, concealing ruptures during testing; fabricated data propagated into customers' qualification records; conduct from 2000 per the adjudication & M1 & Incident-triggered discovery & Wire-fraud plea; USD 1B penalty (25M fine, 975M restitution), monitor; recalls per NHTSA ``the largest and most complex vehicle recalls in U.S. history'' \\
Nissan & 2017 & Japan & \tierswatch{tierAck}~Acknowledged &Completion inspection certificates signed under trainee authority rather than by certified final inspectors, at five plants as normal practice & M3 & Regulator inspection & Recall and re-inspection of approximately 1.21M vehicles; investigation report to MLIT of 17 Nov 2017 \\
Kobe Steel & 2017- 2019 & Japan & \tierswatch{tierAdj}~Adjudicated &Supplier inspection certificates rewritten so non-conforming aluminium and copper shipped as conforming; falsified certificates entered hundreds of customers' records & M4 & Self-report & About 500 customers admitted 2017, more than 600 by the March 2018 report; indictment July 2018; Tachikawa Summary Court fine of 100M yen (March 2019) \\
Subaru & 2017- 2018 & Japan & \tierswatch{tierAck}~Acknowledged &Completion inspections pass/fail decided by staff not appointed as final inspectors; certificate authority did not match the system's requirement & M3 & Self-report & Recall of 371,032 vehicles (16 Nov 2017); further final-inspection recalls of 25,979, 5,879, and 97,001 through Nov 2018 \\
Hino & 2022- 2025 & Japan/ US & \tierswatch{tierAdj}~Adjudicated &Company-disclosed falsification of emissions-durability and fuel-economy certification data (Japan), incl. acknowledged false reporting to MLIT in 2016; adjudicated false US certification applications and altered test data, 2010-2019 & M1 & Self-report & MLIT type designation revocations and correction order (2022); US plea, sentenced 19 Mar 2025: USD 521.76M fine, USD 1.087B forfeiture, five-year probation \\
Daihatsu & 2023- 2024 & Japan & \tierswatch{tierAck}~Acknowledged &False statements in test reports, improperly modified test vehicles, manipulated raw data, per the accepted third-party panel (174 irregularities, 64 models) and MLIT's record (156 cases, 46 models) & M1 & Self-report & All-model shipment halt from 20 Dec 2023; correction order; type designations of three models revoked 26 Jan 2024; staged resumption to 19 Apr 2024; 85-company MLIT order \\
Toyota Industries & 2024 & Japan & \tierswatch{tierAck}~Acknowledged &Diesel output certification measured with non-production ECU software; evidence did not represent the production configuration & M2 & Self-report & Engine and vehicle shipment suspensions from 29 Jan 2024 covering ten models incl. Land Cruiser 300 and Hilux \\
Five makers, 3 June 2024 & 2024 & Japan & \tierswatch{tierAck}~Acknowledged &Test data generated under non-prescribed conditions (impact angles, barrier masses, vehicle weights, ECU software, airbag timers); some values rewritten; two Toyota tests harsher than required & M2 & Self-report (within MLIT-ordered review) & Toyota halted shipments, sales, and production of three models; Mazda shipments of two models remained suspended from 30 May; MLIT on-site inspections of all five with NALTEC verification; disclosures followed MLIT's 26 Jan 2024 order to 85 firms \\
\bottomrule
\end{tabular}
\end{table*}

Table 3 assembles the compendium: thirteen compliance-documentation failures made public 2012-2024, each spanning disclosure to resolution, though the conduct often predates that, to 1989 at Daihatsu. Five are adjudicated and eight company-acknowledged, the alleged category unoccupied: every case rests on an adjudication instrument or the subject's own admission. The sole alleged element, an indictment of three Takata executives, is component-level; Appendix B carries full chronologies and sources.

The nine Japanese cases form one arc, read as the industry reads it: a certification system under growing load, failing at successive points of the evidence chain, surfaced mostly by the companies themselves. \Cref{fig:japantimeline} places this cluster along a 2016-2024 timeline above the axis, the global comparators below it, marker colour keying each case's evidence tier.

In April 2016 Mitsubishi Motors admitted improper fuel-consumption testing by a running-resistance method non-prescribed since 1991. Nissan, the customer, surfaced the deviation; adjudicated Consumer Affairs Agency surcharge orders followed in 2017. \caseinsight{Cell: M1 evidence fabrication at Create/Generate, customer discovery.} Suzuki then disclosed that driving-resistance data for 26 models came from component measurements, not the regulated coasting test. Its MLIT report indicated the catalogue figures were not overstated, citing workload. \caseinsight{Cell: M2 procedural non-equivalence at Create/Generate, self-reported.}

In 2017 the failure moved to sign-off, which Japan's type designation delegates to the manufacturer. Nissan acknowledged that final inspections had routinely been signed by uncertified inspectors, recalling about 1,210,000 vehicles after MLIT inspections. Subaru acknowledged that uncertified inspectors had likewise made pass/fail determinations. \caseinsight{Both cells: M3 unauthorised approval at Approve/Sign; regulator inspection at Nissan, self-report at Subaru.}

Kobe Steel disclosed in October 2017 that inspection-certificate data for aluminium and copper had been improperly rewritten, affecting about 500 customers. \cite{caseKobe} The Tachikawa Summary Court fined it 100 million yen in March 2019 under the Unfair Competition Prevention Act. The falsified artefact was the inter-firm conformity certificate, validated wholly inside the issuer: the EN 10204 type 3.1 pattern that type 3.2's independent countersignature exists to correct. \caseinsight{Cell: M1 at the supplier, M4 upstream propagation at Verify/Reconcile in the chain, self-reported.}

Hino Motors acknowledged in March 2022 falsifying engine certification data over many years, and MLIT revoked the affected type designations that month. \cite{caseHino} The US component is adjudicated: a guilty plea to conspiracy to defraud the US government and consumers, sentenced in March 2025 at over USD 1.6 billion. \caseinsight{Cell: M1 at Create/Generate, self-reported.}

Daihatsu acknowledged its third-party committee's findings, published in December 2023: 174 irregularities across 64 models, which the panel tied to rigid schedules and first-attempt pass pressure. \cite{caseDaihatsu} It halted shipment of all Daihatsu-developed models. MLIT's own narrower count was 156 irregularities across 46 models, and on 26 January 2024 it revoked three models' type designations over modified test vehicles. \caseinsight{Cell: M1 at Create/Generate, volume and deadline pressure, self-reported.}

On 26 January 2024 MLIT also ordered 85 type-designation holders to investigate their applications, reporting by end-May. Three days later Toyota Industries disclosed that horsepower for three diesel engines was measured using ECUs with non-production software; Toyota suspended ten models including the Land Cruiser 300 and Hilux. On 30 January the chairman apologised over Hino, Daihatsu and Toyota Industries. \caseinsight{Cell: M2 at Create/Generate, self-reported.}

On 3 June 2024 five manufacturers, Toyota, Mazda, Yamaha Motor, Honda and Suzuki, reported type-designation irregularities to MLIT. \cite{caseJune2024,caseToyota2024} Toyota disclosed seven models and halted shipments, sales and Japanese production of three. Two instances stand out: pedestrian-protection data at 65 degrees against a stipulated 50, and fuel-leakage data with an 1,800 kg barrier against a specified 1,100 kg. Both were harsher than required, yet procedurally non-compliant. Mazda, Honda, Suzuki and Yamaha reported comparable rewritten values and non-prescribed test conditions, mostly on past models, Mazda's two current models suspended since 30 May. Akio Toyoda said ``We are not a perfect company''. In AP's paraphrase, he said the company may have been too eager as model varieties grew, short deadlines burdening certification. \caseinsight{Event cell: M2 at Create/Generate, volume and deadline pressure, self-reported.}

The remaining cases sit outside the Japanese certification arc. From 2000 Takata induced automaker customers to buy inflators with false test reports concealing test ruptures. \cite{caseTakata} It pleaded guilty to wire fraud in February 2017, an adjudicated resolution, sentenced to USD 1 billion. NHTSA calls the resulting recall of about 46 million inflators the largest in US automotive history. Three executives were separately indicted; those charges remain allegations, the individuals presumed innocent. \caseinsight{Cell: M1 at Create/Generate, incident-triggered discovery; downstream spread a consequence, not a second mechanism.}

Volkswagen is the boundary case, its value in what did not break. EPA's September 2015 notice of violation alleged software that ran full emissions controls only under official testing; Volkswagen then pleaded guilty to three felony counts, an adjudicated outcome, sentenced April 2017. \cite{caseVW} By EPA's account, university researchers raised the questions that prompted EPA and CARB investigation. The certification documents were facially complete and internally consistent, the test data real measurements as presented, and the adjudicated falsity lay in the artefact's engineered unrepresentativeness under test. No documentary cross-check inside the chain could expose it; only observing the artefact outside the test condition could, as the regulators' in-use work did. \caseinsight{Cell: M5 artefact-evidence decoupling at Create/Generate, third-party discovery: document-centric assurance's outer edge.}

GM India recalled 114,000 Chevrolet Taveras in 2013 over emissions compliance. \cite{caseGMIndia} GM acknowledged that an internal investigation found employees setting fine-tuned lower-emission engines aside for inspections. A government committee was reported to have found testing rules flouted. \caseinsight{Cell: M2 at Verify/Reconcile, internal-audit escalation.}

Hyundai and Kia lowered fuel-economy labels for most of their 2012-2013 fleet in November 2012 after EPA testing diverged from submitted data. \cite{caseHyundaiKia} The 2014 consent decree, an adjudicated resolution, brought a USD 100 million civil penalty. \caseinsight{Cell: M2 at Create/Generate, regulator inspection.}

Across the thirteen, detection came by self-report, regulator action, third-party or customer discovery, internal escalation, or incident; on this record, not one failure surfaced through the chain's own routine verification. Section 6 develops that observation.

\section{The failure-mode taxonomy}\label{sec:taxonomy}
\begin{figure*}[t]\centering
\resizebox{\textwidth}{!}{
\begin{tikzpicture}[
  font=\sffamily\small,
  card/.style={draw=sapphire, line width=1pt, rounded corners=4pt, fill=white,
               text width=3.05cm, minimum height=5.4cm, align=left, inner sep=6pt,
               text=inkGrey, anchor=north},
  boundarycard/.style={draw=tierAlg, line width=1.1pt, dashed, rounded corners=4pt,
               fill=tierAlgFill!45, text width=3.05cm, minimum height=5.4cm, align=left,
               inner sep=6pt, text=inkGrey, anchor=north}
]
\newcommand{\midT}[1]{{\sffamily\large\bfseries\color{sapphireDark}#1}}
\newcommand{\midB}[1]{{\sffamily\large\bfseries\color{tierAlg}#1}}
\newcommand{\nmT}[1]{{\sffamily\small\bfseries\color{inkGrey}#1}}
\newcommand{\stgT}[1]{{\sffamily\scriptsize\bfseries\color{sapphireMid}#1}}
\newcommand{\csT}[1]{{\sffamily\scriptsize\itshape\color{inkGrey}#1}}
\newcommand{\hdrT}[1]{{\sffamily\small\bfseries\color{sapphireDark}#1}}

\def\dx{3.45cm}

\node[card] (m1) at (0,0)
  {{\midT{M1}}\\[1pt] {\nmT{Fabrication}}\\[4pt]
   Evidence invented or altered so a result that never occurred appears to pass.\\[6pt]
   {\stgT{Hits: Create / Generate}}\\[5pt]
   {\csT{canonical: fabricated fuel-economy or emissions data}}};

\node[card] (m2) at (\dx,0)
  {{\midT{M2}}\\[1pt] {\nmT{Procedural}}\\ {\nmT{non-equivalence}}\\[4pt]
   The test ran, but by a method the standard does not prescribe.\\[6pt]
   {\stgT{Hits: Create / Generate}}\\[5pt]
   {\csT{canonical: 65\textdegree{} vs 50\textdegree{} pedestrian test}}};

\node[card] (m3) at (2\dx,0)
  {{\midT{M3}}\\[1pt] {\nmT{Unauthorised}}\\ {\nmT{approval}}\\[4pt]
   A required sign-off made by a person or body lacking the certified authority.\\[6pt]
   {\stgT{Hits: Approve / Sign}}\\[5pt]
   {\csT{canonical: uncertified final-inspection technicians}}};

\node[card] (m4) at (3\dx,0)
  {{\midT{M4}}\\[1pt] {\nmT{Upstream}}\\ {\nmT{propagation}}\\[4pt]
   Falsified supplier evidence is accepted unaltered into OEM records at receiving verification.\\[6pt]
   {\stgT{Hits: Verify / Reconcile}}\\[5pt]
   {\csT{canonical: falsified material certificates (Kobe Steel)}}};

\node[boundarycard] (m5) at (4\dx,0)
  {{\midB{M5}}\\[1pt] {\nmT{Artefact--evidence}}\\ {\nmT{decoupling}}~\boundary\\[4pt]
   The artefact attests conformity while the product's real behaviour diverges by design.\\[6pt]
   {\stgT{Hits: Create / Generate}}\\[5pt]
   {\csT{canonical: defeat-device emissions case (Volkswagen)}}};

\node[font=\sffamily\small\bfseries, text=sapphireDark, anchor=south] at ($(m1.north)!0.5!(m5.north)+(0,0.35)$)
  {Five documentation failure modes, mapped to the companion paper's compliance-document lifecycle};
\draw[sapphireMid, line width=0.8pt]
  ($(m1.north)+(0,0.2)$) -- ($(m5.north)+(0,0.2)$);
\foreach \c in {m1,m2,m3,m4,m5}{
  \draw[sapphireMid, line width=0.8pt] ($(\c.north)+(0,0.2)$) -- ($(\c.north)+(0,0.05)$);}

\node[fit=(m1)(m2)(m3)(m4)(m5), inner sep=0pt, draw=none] (cards) {};
\node[font=\sffamily\scriptsize, text=inkGrey, align=left, text width=15.5cm, anchor=north west]
  at ($(cards.south west)+(0,-0.3)$)
  {\normalsize\normalfont
   M1--M4 are in-scope conformity-evidence failures; {\color{tierAlg}\itshape M5 is the boundary case}, where the evidence is technically produced yet the attestation is defeated.};

\end{tikzpicture}}
\caption{The documentation failure-mode taxonomy. Modes M1 to M4 are conformity-evidence failures, each mapped to the compliance-document lifecycle stage it strikes; M5, artefact-evidence decoupling, is the boundary case, where the evidence is technically produced yet the attestation is defeated.}\label{fig:taxonomy}
\end{figure*}
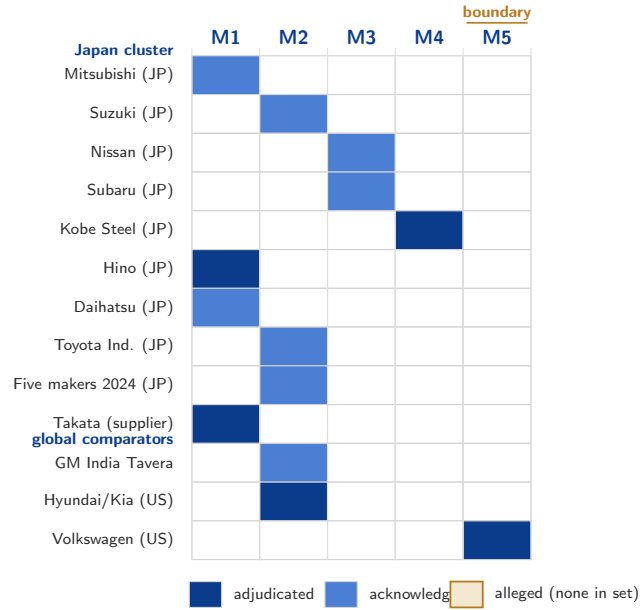
\begin{figure}[t]\centering
\resizebox{\columnwidth}{!}{
\begin{tikzpicture}[
  font=\sffamily\scriptsize,
  x=1.05cm, y=0.60cm,
  colh/.style={font=\sffamily\small\bfseries, text=sapphireDark},
  rowh/.style={font=\sffamily\scriptsize, text=inkGrey, anchor=east},
  cell/.style={draw=gray!35, line width=0.5pt, minimum width=1.05cm, minimum height=0.60cm},
  adj/.style={cell, fill=tierAdj},
  ack/.style={cell, fill=tierAck},
  alg/.style={cell, fill=tierAlgFill, draw=tierAlg, line width=0.8pt},
  empty/.style={cell, fill=white}
]

\foreach \m/\cx in {M1/1, M2/2, M3/3, M4/4, M5/5}{
  \node[colh] at (\cx,1) {\m};}
\node[colh, text=tierAlg] at (5,1.62) {\scriptsize boundary};
\draw[tierAlg, line width=0.7pt] (4.55,1.35) -- (5.45,1.35);

\node[font=\sffamily\scriptsize\bfseries, text=sapphireDark, anchor=east] at (0.35,0.62) {Japan cluster};
\node[font=\sffamily\scriptsize\bfseries, text=sapphireDark, anchor=east] at (0.35,-9.4) {global comparators};

\node[rowh] at (0.35,0)  {Mitsubishi (JP)};
\node[ack] at (1,0){}; \node[empty] at (2,0){}; \node[empty] at (3,0){}; \node[empty] at (4,0){}; \node[empty] at (5,0){};
\node[rowh] at (0.35,-1) {Suzuki (JP)};
\node[empty] at (1,-1){}; \node[ack] at (2,-1){}; \node[empty] at (3,-1){}; \node[empty] at (4,-1){}; \node[empty] at (5,-1){};
\node[rowh] at (0.35,-2) {Nissan (JP)};
\node[empty] at (1,-2){}; \node[empty] at (2,-2){}; \node[ack] at (3,-2){}; \node[empty] at (4,-2){}; \node[empty] at (5,-2){};
\node[rowh] at (0.35,-3) {Subaru (JP)};
\node[empty] at (1,-3){}; \node[empty] at (2,-3){}; \node[ack] at (3,-3){}; \node[empty] at (4,-3){}; \node[empty] at (5,-3){};
\node[rowh] at (0.35,-4) {Kobe Steel (JP)};
\node[empty] at (1,-4){}; \node[empty] at (2,-4){}; \node[empty] at (3,-4){}; \node[adj] at (4,-4){}; \node[empty] at (5,-4){};
\node[rowh] at (0.35,-5) {Hino (JP)};
\node[adj] at (1,-5){}; \node[empty] at (2,-5){}; \node[empty] at (3,-5){}; \node[empty] at (4,-5){}; \node[empty] at (5,-5){};
\node[rowh] at (0.35,-6) {Daihatsu (JP)};
\node[ack] at (1,-6){}; \node[empty] at (2,-6){}; \node[empty] at (3,-6){}; \node[empty] at (4,-6){}; \node[empty] at (5,-6){};
\node[rowh] at (0.35,-7) {Toyota Ind. (JP)};
\node[empty] at (1,-7){}; \node[ack] at (2,-7){}; \node[empty] at (3,-7){}; \node[empty] at (4,-7){}; \node[empty] at (5,-7){};
\node[rowh] at (0.35,-8) {Five makers 2024 (JP)};
\node[empty] at (1,-8){}; \node[ack] at (2,-8){}; \node[empty] at (3,-8){}; \node[empty] at (4,-8){}; \node[empty] at (5,-8){};

\node[rowh] at (0.35,-9) {Takata (supplier)};
\node[adj] at (1,-9){}; \node[empty] at (2,-9){}; \node[empty] at (3,-9){}; \node[empty] at (4,-9){}; \node[empty] at (5,-9){};
\node[rowh] at (0.35,-10){GM India Tavera};
\node[empty] at (1,-10){}; \node[ack] at (2,-10){}; \node[empty] at (3,-10){}; \node[empty] at (4,-10){}; \node[empty] at (5,-10){};
\node[rowh] at (0.35,-11){Hyundai/Kia (US)};
\node[empty] at (1,-11){}; \node[adj] at (2,-11){}; \node[empty] at (3,-11){}; \node[empty] at (4,-11){}; \node[empty] at (5,-11){};
\node[rowh] at (0.35,-12){Volkswagen (US)};
\node[empty] at (1,-12){}; \node[empty] at (2,-12){}; \node[empty] at (3,-12){}; \node[empty] at (4,-12){}; \node[adj] at (5,-12){};

\foreach \cx in {0.5,1.5,2.5,3.5,4.5,5.5}{
  \draw[gray!25, line width=0.4pt] (\cx,0.7) -- (\cx,-12.5);}

\node[adj, minimum width=0.5cm, minimum height=0.4cm] at (0.7,-13.4){};
\node[font=\sffamily\scriptsize, text=inkGrey, anchor=west] at (1.0,-13.4){adjudicated};
\node[ack, minimum width=0.5cm, minimum height=0.4cm] at (2.7,-13.4){};
\node[font=\sffamily\scriptsize, text=inkGrey, anchor=west] at (3.0,-13.4){acknowledged};
\node[alg, minimum width=0.5cm, minimum height=0.4cm] at (4.55,-13.4){};
\node[font=\sffamily\scriptsize, text=inkGrey, anchor=west] at (4.85,-13.4){alleged (none in set)};

\end{tikzpicture}}
\caption{The case-by-failure-mode matrix. A filled cell marks the mode a case is assigned; cell colour keys the case's R5 tier. M5 appears only as the boundary case, isolating the defeat-device pattern from the conformity-evidence failures. Empty cells are unclaimed, not absent.}\label{fig:matrix}
\end{figure}
\begin{table*}[t]
\centering
\caption{The five documentation failure modes: definition, the compliance-document lifecycle stage each strikes, and the cases in the compendium exhibiting it. M5 is the boundary case.}
\label{tab:modes}
\footnotesize
\begin{tabular}{@{}p{0.9cm}p{2.55cm}p{6.5cm}p{2.3cm}p{3.7cm}@{}}
\toprule
\headrow \hcell{Mode} & \hcell{Failure} & \hcell{Definition} & \hcell{Lifecycle stage} & \hcell{Cases in set} \\
\midrule
\mode{M1} & Fabrication & Evidence invented or altered so a non-occurring result appears to pass & Create/Generate & Mitsubishi, Hino, Daihatsu, Takata (4) \\
\mode{M2} & Procedural non-equivalence & Test performed by a method not equivalent to the prescribed one, including harsher-than-required conditions & Create/Generate & Suzuki, Toyota Industries, June 2024, GM India, Hyundai/Kia (5) \\
\mode{M3} & Unauthorised approval & A required sign-off made by a person or body lacking the certified authority & Approve/Sign & Nissan, Subaru (2) \\
\mode{M4} & Upstream propagation & Falsified supplier evidence accepted unaltered into OEM records at receiving verification & Verify/Reconcile & Kobe Steel (1) \\
{\bfseries\color{tierAlg}M5} & Artefact-evidence decoupling (boundary) & The artefact attests conformity while the product's behaviour diverges by design & Create/Generate & Volkswagen (1) \\
\bottomrule
\end{tabular}
\end{table*}

The taxonomy classifies every compendium case along four dimensions. Dimension one, mechanism, what broke in the evidence. M1, evidence fabrication: test data or records invented or altered. M2, procedural non-equivalence: results real, procedure non-prescribed. M3, unauthorised approval: sign-off by persons or under authority not permitted. M4, upstream propagation: falsified supplier evidence flowing into many customers' compliant-looking records. M5, artefact-evidence decoupling: documentation internally consistent while the artefact is engineered to behave unrepresentatively under test. \Cref{fig:taxonomy} presents these five mechanisms as reference cards, and Table 4 states each definition. Dimension two, lifecycle locus, the stage at which the failure entered the record; its five stages (Create/Generate, Verify/Reconcile, Approve/Sign, Record/Retain, Retrieve/Reconstruct) adopt the companion manuscript's lifecycle model, not re-derived here \xcitePone. Dimension three, driver, at most two per case: volume and deadline pressure, incentive misalignment, specification ambiguity, weak binding of approval to artefact and authority, or concealment intent. Dimension four, detection path: self-report, whistleblower or internal-audit escalation, regulator inspection, incident-triggered discovery, or third-party or customer discovery.

Two rules govern the assignments. First, case individuation: one case per distinct disclosure event by a responsible organisation. The 3 June 2024 disclosures therefore form one case, a single MLIT-announced event covering five makers. Mitsubishi and Suzuki in 2016, by contrast, entered as separate disclosures with distinct conduct, so they stand as two cases (Section 5). Second, dominant mechanism: a composite case takes the mechanism of its most prevalent component. The June 2024 case is assigned M2 because non-prescribed test conditions dominate the disclosed instances, including both harsher-than-required Toyota instances. The values rewritten or falsely stated at Honda, Suzuki, and Yamaha meet M1's definition and are acknowledged as M1 elements inside the composite.

\Cref{fig:matrix} plots the thirteen cases against the five mechanisms; Table 3 lists each case's mechanism and detection path. Exhaustiveness holds under these rules: every case received exactly one mechanism, one locus, one detection path, and one or two drivers; none resisted assignment, and the reserved extra mechanism was not needed. Occupancy is M1 four cases, M2 five, M3 two, M4 one, M5 one, so no mechanism is merely theoretical.

Discrimination, that no two mechanisms collapse, is argued at the boundaries. First, M1 against M2, fixed by the 2016 pair, both running-resistance data before the same regulator (Section 5). Mitsubishi's recorded values presented fuel economy better than actual: the content was misrepresentative, hence M1, the long-standing non-prescribed method secondary. Suzuki instead used component-measurement data rather than the regulated coasting test, and its own re-tests indicated the figures were not overstated: procedure broke while content held, M2. Same artefact type, different mechanism: the cut is real, not nominal.

Second, M2 against M5. Toyota Industries measured real engines under a non-production software configuration: the procedure deviated, but the tested artefact was the artefact sold. Volkswagen was convicted of circumventing the US test with software that detected the official emissions test and enabled full controls only for its duration; the documentation was internally consistent while the artefact itself was unrepresentative, which no procedural check could expose. GM India sits with M2: prepared engines at the verification event (Section 5) are a unit-selection deviation a specification can control, at Verify/Reconcile, not M5.

Third, M1 against M4: the cut turns on the falsified artefact's role in the chain, not the detection path. Kobe Steel is M4 because the falsified artefact was the cross-party conformity certificate itself, consumed by downstream firms' receiving verification; that certificate-acceptance step is the defining failure, at Verify/Reconcile. Takata is M1 because the falsified artefact was component test evidence supplied to its direct customers: fabrication at creation is the defining break, though propagation followed. Fourth, M3 is not a data mechanism: the Nissan and Subaru inspections took place and their content is not impugned (Section 5). What broke is approval provenance; no other mechanism describes it, and M3 describes nothing else.

O1: routine verification detected nothing. None of the thirteen cases surfaced through the routine document-facing layer of Section 4: submissions dispositioned, certificates accepted, audits passed. Detection came instead from self-report in seven, regulator enforcement in two, third-party or customer discovery in two, and field incident and internal escalation once each. The verification layer appears in the record only as the broken stage (M3, M4), never as the discovering party.

O2: corruption enters upstream. Nine of thirteen cases sit at Create/Generate; two at Verify/Reconcile, two at Approve/Sign. Once evidence exists, or its creation conditions deviate, every downstream stage transmits it.

O3: the largest mechanism, M2, includes testing harsher than required. Toyota submitted pedestrian-protection development data at a 65-degree impact angle where the regulation stipulated 50, more severe by the company's own account yet non-compliant. In rear-collision fuel-leakage tests, development data from an 1,800 kg moving barrier were submitted where 1,100 kg was specified. A regime that cannot distinguish over-testing from under-testing checks procedural conformity, not risk; this motivates Section 8's machine-readable procedure requirement.

The empty Record/Retain and Retrieve/Reconstruct columns are a visibility statement about the public record, not a safety statement: retention and retrieval failures surface as audit findings and recall-execution friction, not national news, so a public-record compendium cannot contain them. The same censoring reaches the supplier-to-OEM segment: its supplier-side members, Takata and Kobe Steel, reached the record through incident and court, while routine documentation failures caught at receiving surface, if at all, in private customer audits. The compendium therefore samples the catastrophic tail of the segment Section 4 systematises, which Section 9 turns into a failure-telemetry agenda. The driver specification ambiguity is likewise unoccupied: the harsher-than-required instances were clear prescriptions breached by capacity and configuration outcomes, not ambiguity, so the driver stays theoretically motivated.

\section{Reading the record through the burden lens}\label{sec:lens}
The taxonomy's sharpest finding is O1: across thirteen cases, the chain's routine document-facing verification layer detected nothing. A chain this dense in checking stages produced not one discovery through them; detection came from self-report, regulator enforcement, field incidents, third-party or customer discovery, and internal escalation. That demands an explanation, one that also says where the next failures will sit.

The companion manuscript supplies the lens, which this paper imports rather than re-argues \xcitePone. Compliance documentation is mandatory, output-adjacent work: the product cannot ship without it, yet none of it shows in the product, and the lens holds such invisible work is systematically under-resourced when capacity tightens. The companion paper names this documentation exergy destruction, borrowing the thermodynamic notion of work potential destroyed wherever a process runs irreversibly \cite{Bejan2016}.

The industry has articulated the same diagnosis in its own words. AP reported, in indirect speech, that Akio Toyoda said the company may have been too eager to get the tests done, abbreviating them at a time when model varieties were burgeoning. In the transcript he acknowledged that workload had increased and that short deadlines and repeated re-dos burdened the final certification stage. He also suggested some certification rules might be overly stringent, while saying he was not condoning the violations. Read through the lens, these remarks describe a verification layer whose load grew with model proliferation while its resourcing did not; the diagnosis is the industry's own, not this paper's imposition.

The lens yields one prediction the record can be held against: failures should cluster where verification burden is highest and organisational visibility lowest. On this paper's reading, the grid is consistent with that prediction. Nine of thirteen cases enter at Create/Generate, the highest-volume and least visible stage of evidence production. The largest mechanism cell is procedural non-equivalence, deviation in how mandated work was performed rather than in what was reported. The burden reading rests on the companies' own stated causes and on the size of that M2 cell, not on a driver tally. Suzuki's report cited workload and under-investment in coasting-test infrastructure; the Daihatsu panel attributed the irregularities to rigid schedules and first-attempt pass pressure; the Toyoda remarks describe burden at the final certification stage. In the driver tally itself, volume and deadline pressure ties with incentive misalignment and concealment intent, at three Japanese placements each. Thirteen cases support a reading, not a statistic; the measurement question, burden and failure rates per unit of verification load, stays open in Section 9.

This paper's contribution to the lens is the failure record it predicts.

\section{Implications for evidence architecture}\label{sec:implications}
Each mechanism in Section 6 names the property whose absence it exploited. This section states the requirements as neutral engineering: properties an evidence architecture must exhibit, independent of implementation, product, or supplier.

M1, evidence fabrication, succeeded because the record was the only witness to the event it described. Two properties counter it. First, independent verification paths: for load-bearing evidence, a second route to the same quantity not passing through the original author, by instrument capture, duplicate measurement, or reconciliation against coupled quantities. Second, tamper-evidence: alteration of a record after creation must leave a detectable trace, so that the cheapest fabrication is no longer edit-in-place. Neither stops a determined fabricator; both raise its cost and make later reconstruction possible.

M2, procedural non-equivalence, is invisible in results: its cases produced real data whose defect lay in how they were produced. The requirement is a machine-readable specification of the prescribed procedure, covering test conditions, equipment and software configuration, and unit-selection rules, against which a test's conduct, not just its output, can be checked before submission. Over-testing and under-testing then become equally visible, a distinction O3 shows the regime does not make.

M3, unauthorised approval, is a provenance break: the approval existed, the authority behind it did not. The requirement is binding of approval to artefact and to authority, systematic at minimum, cryptographic where the chain crosses trust boundaries. Every sign-off must carry who approved, what (identity and revision), and under which delegated authority; one outside that authority must be unrecordable or immediately visible.

M4, upstream propagation, exploited downstream acceptance that rested on the issuer's own signature. The materials world already states the counter-principle in certificate grades. An EN 10204 type 3.1 certificate is validated by the manufacturer's authorised inspection representative, independent of the manufacturing department but internal to the manufacturer. A type 3.2 certificate additionally requires validation and countersignature by a party independent of the manufacturer, either the purchaser's authorised representative or an independent third-party inspection body. The requirement generalises the 3.2 independence principle: any certificate whose acceptance feeds many downstream compliance records must be verifiable across parties, so acceptance rests on issuer-independent confirmation, not the issuer's signature alone.

M5, artefact-evidence decoupling, defines what no evidence architecture can promise. Volkswagen's records were internally consistent; the artefact carried software that detected official testing and ran full emissions controls only during it, and the company was convicted of circumventing the US testing process by these means. No strengthening of the preceding requirements would have flagged those records, because nothing in them was wrong; the failure showed only in the artefact's behaviour outside the test. The requirement is honesty of scope: an evidence architecture must state that document-centric assurance cannot alone guarantee artefact behaviour, and must name where its assurance ends and in-use measurement begins.

These are requirements, not designs. Whether process, tooling, or regulation meets them is an implementation question beyond this paper's scope; Section 9 records what must be known before any can be evaluated.

\section{Open problems and research agenda}\label{sec:open}
Five problems follow directly from the record.

(a) Failure telemetry. No public, structured dataset of compliance-documentation failures exists; this paper's compendium was assembled case by case from adjudications, disclosures and regulator records. Table 3's columns (case, years, country, category, what broke, mechanism, detection path, consequence), each bound to a source instrument, plus the locus and driver assignments in Section 6 and \Cref{fig:matrix}, form the seed schema this agenda proposes. The harder problem is censoring: the record over-samples failures spectacular enough to reach adjudication or national news, so the lifecycle's retention and retrieval stages are unobserved here, not necessarily sound. Their failures surface as audit findings and recall-execution friction, channels this record cannot see. A telemetry corpus from audit and recall records would de-censor those stages and let the taxonomy's empty cells be tested, not assumed.

(b) Procedure-as-code for certification tests. The largest mechanism cell in Section 6 is procedural non-equivalence: results real, procedure non-prescribed. Detecting it today requires reconstructing what a test laboratory actually did, against prose regulation. Representing prescribed procedure machine-readably, so conduct logs can be checked for equivalence before submission, would convert the dominant failure mechanism into a detectable error. Japan's amended rules require post-designation sampling tests on actual production vehicles from 1 April 2026; such tests verify the artefact, while a procedure-as-code layer would verify the conduct, the two complementary.

(c) Approval provenance across firm boundaries. The unauthorised-approval and upstream-propagation mechanisms both break at the seam between organisations, where one party's sign-off becomes another party's accepted evidence. The open infrastructure question is what an approval-to-artefact-to-authority binding looks like across a firm boundary. Convergence in the document standards makes the question more tractable: the 7th edition of VDA Volume 2 (January 2026) adopts a risk-based scope and records the German industry's openness to a common international standard for production process and product approval in the long term.

(d) The regime question. Does self-certification produce a measurably different failure distribution from type approval? The type-approval cases surfaced through introspection, largely self-reports within ordered reviews. The two US emissions cases sit under the EPA certificate-of-conformity regime; they surfaced through regulator and third-party testing. The one US safety case, Takata, sits instead under FMVSS self-certification enforced post-hoc by NHTSA, and surfaced through a field incident. Thirteen cases make this suggestive, not sufficient. The regimes are themselves moving: Japan's reforms are fully in force, and Euro 7 applies to new EU types from November 2026, so the coming record forms under changed detection conditions, a natural experiment worth instrumenting now.

(e) Burden measurement. The burden lens of Section 7 lacks a denominator: no one publishes documents produced, verification hours spent, or approvals executed per vehicle programme. Without that denominator, failure rates cannot be normalised, the cost of any Section 8 requirement cannot be estimated, and the industry's own overload diagnosis cannot be tested. Measuring it is the least glamorous problem here, and the one on which the other four depend.

\section{Conclusion}\label{sec:conclusion}
The automotive supplier-quality and certification evidence chain is operationally central and analytically invisible. This paper has mapped it. It systematised the chain across AIAG PPAP and VDA Volume 2 PPA, organised by artefact, producer, verifier, trigger, approval state, and retention obligation, within APQP and IATF 16949 (C1). It then assembled a verified compendium of thirteen documented compliance-documentation failures made public 2012-2024, each classified as adjudicated, company-acknowledged, or alleged, and built strictly from the public record (C2). Every case was placed in a failure-mode taxonomy dimensioned by mechanism, lifecycle locus, driver, and detection path, and the taxonomy shown exhaustive and discriminating over the case set (C3). Finally, it derived, mechanism by mechanism, the evidence-architecture property that would counter each failure, and set an open-problems agenda (C4).

One finding stands above the rest. Across all thirteen cases, not one failure surfaced through the chain's own routine verification; the record indicts the verification layer, not only the evidence authors. The chain is dense in artefacts and thin in checks of its own operation. Until the verification layer receives the engineering attention the evidence layer has long absorbed, this failure record should be read as a baseline, not a closed chapter.

\FloatBarrier

\section*{Competing interests}
The authors are affiliated with Erd\H{o}s Systems, a company developing software for compliance-document processing in the automotive sector; Dawar Jyoti Deka is its founder. This work received no external funding.

\section*{Acknowledgements}
[PLACEHOLDER: pending consent; see FLAGS.md]

\label{refsstart}
\bibliographystyle{plainnat}
\bibliography{refs}

\appendix
\onecolumn

\noindent\emph{Note: Appendix B lists source titles in English; original-language (Japanese) titles, URLs, tiers and access dates are carried verbatim in the claims ledger and \texttt{state/traceability.csv}. Each source group is labelled with the ledger claim ID it supports.}\par\medskip

\section{The PPAP evidence package}\label{app:ppap}
The AIAG PPAP evidence package comprises 18 elements. Table A1 characterises all 18: thirteen are consistently named in the public summaries on which this paper relies, and the remaining five are carried by both independent explainers fetched for this appendix. The AIAG PPAP manual remains the normative reference for the enumeration. In keeping with the copyright discipline applied throughout, purposes are described in our own words and no manual text is reproduced. The producer is in every case the supplier organisation, assembling the package as an output of APQP work. The verifier of record is the customer, whose disposition is captured on the Part Submission Warrant. The final column marks whether the stated obligation is general practice under the base manual and IATF 16949, or arises from customer-specific requirements (CSR).

\smallskip\noindent Table A1. The PPAP evidence package by element.

{\footnotesize\centering
\begin{longtable}{p{2.1cm}p{5.0cm}p{2.4cm}p{2.6cm}p{1.7cm}}
\toprule
Element & Purpose & Producer & Typical verifier & Basis \\
\midrule\endhead
Design records & The released design description whose requirements the process must be shown to meet & Supplier, compiling the applicable design source & Customer quality, at disposition & General \\
Engineering change documentation & Authorised engineering change documents, recording approved changes where a change triggered the submission & Supplier & Customer quality & General \\
Customer engineering approval & Evidence of approval by the customer's engineering function, when required & Supplier, holding the approval evidence & Customer engineering & General \\
Design FMEA & Structured risk analysis of design failure modes, since June 2019 under the harmonised AIAG and VDA seven-step method & Supplier engineering, cross-functional & Customer review at submission & General \\
Process flow diagram & Describes the manufacturing sequence that the process FMEA and control plan then analyse and control & Supplier & Customer quality & General \\
Process FMEA & Structured risk analysis of process failure modes, feeding the control plan; harmonised method as above & Supplier, cross-functional & Customer review at submission & General \\
Control plan & The planned controls (measurements, inspections, monitored process parameters) per product and process step; a living document updated when product, process, equipment, measurement system, or requirement changes & Supplier & Customer at PPAP; auditors thereafter & General; AIAG CP-1 (2024) is the normative reference \\
MSA studies & Statistical evidence, typically Gage R\&R, of the variation of each inspection, measurement, and test equipment system in the control plan, per IATF 16949 clause 7.1.5.1.1 & Supplier quality & Customer at submission & General; IATF-required \\
Dimensional results & Measured results covering the ballooned drawing characteristics; incomplete coverage is a commonly reported rejection cause & Supplier quality & Customer quality, at disposition & General \\
Records of material and performance test results & Documented results of the material and performance tests performed on the part, including pass or fail outcomes; the element through which EN 10204-type material certificates enter the package (Sections 4 and 5) & Supplier quality & Customer quality, at disposition & General \\
Initial process studies & Capability evidence against the customary 1.67 index & Supplier quality & Customer quality & General: the AIAG PPAP manual carries its own initial-study acceptance criterion, which CSRs restate and harden; Ford requires Ppk above 1.67 for initial and final capability per its CSR \\
Qualified laboratory documentation & Industry certifications of the laboratories used for validation testing & Supplier, compiling laboratory credentials & Customer quality & General \\
Appearance approval report & Approval evidence for appearance requirements & Supplier & Customer quality & General \\
Sample production parts & Sample parts from production, provided to the customer for approval, with a record of their storage & Supplier & Customer quality & General \\
Master sample & A retained reference part & Supplier & Customer quality & General \\
Checking aids & The part-specific aids used in checking the product & Supplier & Customer quality & General \\
Customer-specific requirements & Evidence of compliance with the customer's CSRs, which can extend every row above; GM, for example, mandates compliance with the AIAG PPAP manual and adds its own clauses & Supplier & Customer & CSR by definition \\
Part Submission Warrant (PSW) & The package's summary and gating sign-off: reason for submission, authorised supplier signature, and the customer's disposition field; required at every submission level & Supplier authorised signatory & Customer disposition on the warrant & General \\
\bottomrule
\end{longtable}
}

A bulk material requirements checklist is not carried as a standalone element by either fetched explainer: bulk material appears only within the customer-specific-requirements element. Two post-approval obligations attach to the package. Annual layout inspection and revalidation are CSR-imposed: GM requires an annual complete layout inspection of all product dimensions on at least five parts throughout the product's life, and Ford requires the annual layout requirement to be included in the control plan. Production part approval records are retained for the period the part is active for production and service plus one calendar year, unless the customer or a regulatory agency specifies longer; GM delegates retention periods to GMW15920.

\section{Case chronologies and sources}\label{app:cases}
This appendix condenses each compendium case into a dated chronology and lists the ledger claims on which Section 5 and Table 3 rest. For each claim, source titles and publishers are given; the claims ledger carries the URLs, source tiers, access dates and anchor quotes. Evidentiary categories and taxonomy cells follow Section 5.

\smallskip\noindent\textbf{B.1 Mitsubishi Motors, 2016--2017 (company-acknowledged; M1, Create/Generate)}\par

\begin{itemize}\setlength{\itemsep}{1pt}
\item \textbf{1991.} Per MMC's later report to MLIT, the company began testing Japan-market vehicles with a non-prescribed high-speed coasting method when the coasting test was designated by regulation.
\item \textbf{20 April 2016.} MMC admitted improper fuel-consumption testing on four minicar models, about 625,000 vehicles including units built for Nissan; production and sales stopped.
\item \textbf{20 April 2016.} MMC's release recorded that Nissan had surfaced the deviations during joint development and asked MMC to review its running-resistance values.
\item \textbf{26 April 2016.} MMC's report to MLIT stated the 1991 origin of the non-prescribed method.
\item \textbf{18 May 2016.} MLIT instructed all manufacturers to investigate their exhaust-emission and fuel-consumption testing.
\item \textbf{27 January 2017.} Consumer Affairs Agency corrective orders and a surcharge payment order totalling 485.07 million yen, the first under the Act's 2016 surcharge system.
\item \textbf{14 June 2017.} Follow-on surcharge orders added 4.53 million yen for MMC and 3.17 million yen for Nissan.
\end{itemize}

\noindent\textbf{Sources.} C-0088: ``Improper conduct in fuel consumption testing on products manufactured by Mitsubishi Motors Corporation (MMC)'' and ``Regarding the Report to MLIT Concerning Improper Conduct in Fuel Consumption Testing of Vehicles Manufactured by Mitsubishi Motors Corporation'', Mitsubishi Motors Corporation (Wayback Machine snapshots). C-0089: ``Report to the Ministry of Land, Infrastructure, Transport and Tourism on the instruction for investigation of inappropriate exhaust-emission and fuel-consumption testing'', Suzuki Motor Corporation. C-0102: ``Operation of the surcharge system under the Premiums and Representations Act'', Consumer Affairs Agency / Cabinet Office, Japan (in Japanese); report on the Consumer Affairs Agency surcharge order against Mitsubishi Motors for fuel-economy misrepresentation, Response.jp (in Japanese).

\smallskip\noindent\textbf{B.2 Suzuki, 2016 (company-acknowledged; M2, Create/Generate)}\par

\begin{itemize}\setlength{\itemsep}{1pt}
\item \textbf{20 April 2016.} Mitsubishi's admission opened the 2016 fuel-economy testing episode.
\item \textbf{18 May 2016.} MLIT instructed all manufacturers to investigate their testing.
\item \textbf{May 2016.} Suzuki disclosed that driving resistance for Japan-market type approval had been assembled from component measurements rather than the regulated coasting test.
\item \textbf{31 May 2016.} Suzuki's report to MLIT identified 26 affected models, found no intention to manipulate fuel-consumption figures, and reported re-tests confirming the catalogue figures were not overstated.
\item \textbf{Stated causes.} Post-2008 workload and under-investment in coasting-test infrastructure.
\end{itemize}

\noindent\textbf{Sources.} C-0088: as in B.1. C-0089: ``Report to the Ministry of Land, Infrastructure, Transport and Tourism on the instruction for investigation of inappropriate exhaust-emission and fuel-consumption testing'', Suzuki Motor Corporation.

\smallskip\noindent\textbf{B.3 Nissan, 2017 (company-acknowledged; M3, Approve/Sign)}\par

\begin{itemize}\setlength{\itemsep}{1pt}
\item \textbf{System context.} Japan's type designation system delegates the per-vehicle conformity check to the manufacturer, which issues a completion inspection certificate under Road Transport Vehicle Act Article 75(4).
\item \textbf{From 18 September 2017.} MLIT conducted on-site inspections; Nissan's later report documents improper responses displayed during them.
\item \textbf{2 October 2017.} Nissan announced non-compliant final inspections; approximately 1,210,000 units produced October 2014 to September 2017 became subject to recall and re-inspection, plus about 34,000 unregistered vehicles.
\item \textbf{17 November 2017.} Nissan's investigation report to MLIT found that at five plants it had become normal practice for final inspectors in training to conduct final-inspection tasks.
\end{itemize}

\noindent\textbf{Sources.} C-0009: ``Interim summary on the direction of improvement and rationalisation of completion inspection in the type designation system'' (21 April 2020), MLIT (in Japanese). C-0084: ``Nissan to re-inspect and resume registrations of vehicles in inventory in Japan, and re-inspect registered vehicles'' and ``Report summary of final vehicle inspection issue at plants in Japan'', Nissan Motor Co., Ltd.

\smallskip\noindent\textbf{B.4 Subaru, 2017--2018 (company-acknowledged; M3, Approve/Sign)}\par

\begin{itemize}\setlength{\itemsep}{1pt}
\item \textbf{System context.} As B.3: the completion inspection certificate is the delegated per-vehicle conformity sign-off.
\item \textbf{October 2017.} Subaru acknowledged improper conduct in final vehicle inspections at its Gunma plants.
\item \textbf{27 October 2017.} Reuters reported uncertified technicians conducting final inspections, a practice Kyodo reported had continued for over 30 years (reported, not company-confirmed).
\item \textbf{16 November 2017.} Subaru filed a domestic recall of 371,032 vehicles across 11 nameplates because inspectors not appointed as final inspectors had made pass/fail determinations.
\item \textbf{February to November 2018.} Follow-up final-inspection recalls added 25,979, 5,879 and 97,001 vehicles.
\end{itemize}

\noindent\textbf{Sources.} C-0009: as in B.3. C-0084: as in B.3 (the Nissan comparator for the M3 pair). C-0085: ``About the recall related to final vehicle inspection'', SUBARU Corporation (in Japanese); ``Subaru Reportedly Allowed Uncertified Technicians to Inspect Vehicles'', Fortune (Reuters).

\smallskip\noindent\textbf{B.5 Kobe Steel, 2017--2019 (adjudicated; M4, Verify/Reconcile)}\par

\begin{itemize}\setlength{\itemsep}{1pt}
\item \textbf{September 2016 to August 2017.} Shipping window of the disclosed products: about 19,300 tonnes of aluminium flat-rolled products and extrusions, 2,200 tonnes of copper products, and 19,400 castings and forgings.
\item \textbf{8 October 2017.} Kobe Steel disclosed that inspection-certificate data had been improperly rewritten, products shipped as meeting agreed specifications.
\item \textbf{Late 2017.} Admissions reported to cover about 500 customers.
\item \textbf{March 2018.} Reuters reported more than 600 affected customers in total following the independent investigation; the company attributed roughly 408 to the aluminium and copper business.
\item \textbf{19 July 2018.} Indictment by the Tokyo District Public Prosecutors Office under the Unfair Competition Prevention Act.
\item \textbf{13 March 2019.} The Tachikawa Summary Court fined Kobe Steel 100 million yen.
\item \textbf{As of 3 July 2026.} No appeal or new proceedings found.
\item \textbf{Architecture note.} EN 10204 type 3.1 certificates are validated inside the manufacturer; type 3.2 adds an independent countersignature, the gap this case exposes.
\end{itemize}

\noindent\textbf{Sources.} C-0086: ``Improper conduct concerning a portion of the aluminum and copper products manufactured by Kobe Steel'' and ``Key Facts Revealed by the Independent Investigation Committee's Investigation'', Kobe Steel, Ltd.; also ``Kobe Steel CEO to Resign Over Data Fraud Scandal as Report Finds More Cases'', Insurance Journal (Reuters). C-0087: ``About the decision on Kobe Steel's violation of the Unfair Competition Prevention Act'', Kobe Steel, Ltd. C-0042: ``4 Types of Inspection Certificates (2.1, 2.2, 3.1, and 3.2) as per EN 10204'', HardHat Engineer; ``Mill Test Certificate (MTC): EN 10204 3.1 vs 3.2'', Projectmaterials; ``What is the Difference Between EN 10204 3.1 and 3.2 Inspection Certificates?'', Holland Applied Technologies. C-0043: the Holland Applied Technologies and Projectmaterials items above; ``EN 10204 Certificate Types Explained: Type 2.1, 2.2, 3.1, and 3.2'', MTR.AI. C-0168: currency-sweep claim; sources listed in B.6.

\smallskip\noindent\textbf{B.6 Hino Motors, 2022--2025 (adjudicated US component; Japanese component company-acknowledged; M1, Create/Generate)}\par

\begin{itemize}\setlength{\itemsep}{1pt}
\item \textbf{4 March 2022.} Hino acknowledged falsified engine performance data in an emissions durability test (A05C) and falsified fuel-economy measurement (A09C, E13C); sales of the three engines and corresponding vehicles suspended.
\item \textbf{29 March 2022.} MLIT administrative dispositions, after hearings on 25 March, revoking the type designations tied to the non-compliant engines.
\item \textbf{1 August 2022.} Special Investigation Committee reported long-term misconduct: emissions misconduct from the 2003 (E6) regulation era, fuel-economy misconduct from 2005 (E7), false reporting to MLIT in 2016.
\item \textbf{9 September 2022.} MLIT correction order and cancellation proceedings for four further engine models.
\item \textbf{15 January 2025.} DOJ and EPA announced Hino's agreement to plead guilty to a one-count criminal information.
\item \textbf{19 March 2025.} Sentencing in the Eastern District of Michigan: USD 521.76 million criminal fine, USD 1.087 billion forfeiture, five years' probation with a diesel-engine import bar; over 105,000 non-conforming engines imported and sold 2010--2022.
\item \textbf{As of 3 July 2026.} No further adjudications found.
\end{itemize}

\noindent\textbf{Sources.} C-0082: ``Misconduct concerning Engine Certification'' and ``Investigation Results by the Special Investigation Committee, and Recurrence Prevention Measures and Other Responses'', Hino Motors, Ltd.; ``[Translation] Investigation Report (Summary), August 1, 2022'', Special Investigation Committee / Hino Motors, Ltd. C-0083: ``Court Sentences Hino Motors Ltd., a Toyota Subsidiary, and Imposes Over \$1.6B in Penalties for Emissions Fraud Scheme'', US Department of Justice; ``Hino Motors, a Toyota Subsidiary, Agrees to Plead Guilty and Pay Over \$1.6B to Resolve Emissions Fraud Scheme'', US Environmental Protection Agency. C-0101: ``Administrative disposition against a vehicle manufacturer'' and ``Response regarding Hino Motors'', MLIT (in Japanese); ``Administrative Action by the Ministry of Land, Infrastructure, Transport and Tourism'', Hino Motors, Ltd.; report on Hino's first type-designation revocation over fuel-economy and emissions test fraud, Response.jp (in Japanese). C-0169: ``Court Sentences Hino Motors Ltd., a Toyota Subsidiary, and Imposes Over \$1.6B in Penalties for Emissions Fraud Scheme'', US EPA. C-0168: ``Development of laws and regulations addressing irregularities in vehicle type designation applications'', MLIT (in Japanese); the US EPA Hino sentencing release above; ``Regulation (EU) 2024/1257 (Euro 7)'', EUR-Lex; ``Federal Motor Vehicle Safety Standards; Modernization of FMVSS No.\ 102 To Accommodate ADS-Equipped Vehicles'', Federal Register (NHTSA).

\smallskip\noindent\textbf{B.7 Daihatsu, 2023--2024 (company-acknowledged; M1, Create/Generate)}\par

\begin{itemize}\setlength{\itemsep}{1pt}
\item \textbf{20 December 2023.} Independent Third-Party Committee report: 174 irregularities in 25 test items across 64 models and 3 engines, oldest dating to 1989; causes attributed to rigid development schedules and first-attempt pass pressure.
\item \textbf{20 December 2023.} Daihatsu suspended shipment of all Daihatsu-developed models in production, in Japan and overseas; MLIT instructed the domestic halt be held until conformity was confirmed.
\item \textbf{21 December 2023 to 9 January 2024.} MLIT on-site inspection confirmed the reported 142 irregularities and found 14 more, an administrative total of 156 across 46 models.
\item \textbf{16 January 2024.} Ministerial correction order; recalls directed for two models.
\item \textbf{26 January 2024.} MLIT revoked the type designations of the Daihatsu Gran Max, Toyota Town Ace and Mazda Bongo (truck types), after starting procedures on 16 January.
\item \textbf{26 January 2024.} MLIT instructed 85 type-designation holders to investigate their own applications and report.
\item \textbf{19 January to 19 April 2024.} Shipment-halt instruction lifted in stages as MLIT verified conformity model by model.
\item \textbf{30 January 2024.} The Toyota chairman apologised for the successive irregularities at Hino, Daihatsu and Toyota Industries.
\end{itemize}

\noindent\textbf{Sources.} C-0076: ``Results of the Investigation by the Independent Third-Party Committee and Future Actions'', Daihatsu Motor Co., Ltd.; ``Investigation report of the Independent Third-Party Committee (summary version)'', Independent Third-Party Committee via Daihatsu (in Japanese); ``Toyota's Daihatsu to Halt All Vehicle Shipments, in Widening Safety Scandal'', Reuters (via Claims Journal). C-0077: the Daihatsu release above; ``Irregularities in Daihatsu's type designation applications: chronology and response list'', MLIT (in Japanese). C-0078, C-0080, C-0093, C-0094: ``MLIT's response to the Daihatsu irregularities'' and the MLIT chronology page above, MLIT (in Japanese); for C-0078 also the Reuters (via Claims Journal) item above. C-0079: the MLIT chronology page above. C-0081: ``Results of manufacturers' investigation reports on irregularities in type designation applications'', MLIT (in Japanese); ``Discovery of inappropriate cases in four-wheel-vehicle type designation applications'', Honda Motor Co., Ltd.\ (in Japanese). C-0073: ``Toyota confirms it's world's top car seller, apologizes for safety scandals'', Fortune; ``Toyota chief apologises for scandals as group vehicle sales hit record'', TechXplore (AFP).

\smallskip\noindent\textbf{B.8 Toyota Industries, 2024 (company-acknowledged; M2, Create/Generate)}\par

\begin{itemize}\setlength{\itemsep}{1pt}
\item \textbf{29 January 2024.} TICO received its special investigation committee's report: horsepower output for three commissioned diesel engine models measured using ECUs with non-production software, results appearing smoother; TICO suspended engine shipments and Toyota suspended the affected vehicles.
\item \textbf{29 January 2024.} Ten affected models globally, six in Japan, including the Land Cruiser 300 and Hilux.
\item \textbf{30 January 2024.} Group-level apology by the Toyota chairman covering Hino, Daihatsu and Toyota Industries.
\end{itemize}

\noindent\textbf{Sources.} C-0074: ``Certification Irregularities at Toyota Industries'', Toyota Motor Corporation; ``Toyota confirms it's world's top car seller, apologizes for safety scandals'', Fortune. C-0075: ``Certification Irregularities at Toyota Industries'', Toyota Motor Corporation. C-0073: as in B.7.

\smallskip\noindent\textbf{B.9 The 3 June 2024 disclosures, 2024 (company-acknowledged; M2, Create/Generate)}\par

\begin{itemize}\setlength{\itemsep}{1pt}
\item \textbf{26 January 2024.} MLIT ordered 85 type-designation holders to investigate and report by end-May.
\item \textbf{30 May 2024.} Mazda suspended shipments of two production models and reported to MLIT.
\item \textbf{3 June 2024.} MLIT announced that five manufacturers (Toyota, Mazda, Yamaha Motor, Honda, Suzuki) had reported irregularities in their type designation applications.
\item \textbf{3 June 2024.} Toyota disclosed seven models; shipments, sales and Japanese production of the Corolla Fielder, Corolla Axio and Yaris Cross halted.
\item \textbf{Toyota detail.} Pedestrian-protection data taken at a 65-degree impact angle where regulation stipulated 50 degrees; rear-collision fuel-leakage data taken with an 1,800 kg barrier where regulation specified 1,100 kg.
\item \textbf{Other makers.} Mazda: non-production engine-control software and timer-activated airbags; Honda: noise-test weights and rewritten output and torque values on past models; Suzuki: brake-test false statements on one past model; Yamaha: improper noise-test conditions and horn-test false statements.
\item \textbf{3 June 2024.} MLIT held shipments of affected current-production models and began on-site inspections of all five companies.
\item \textbf{Press conference.} Toyoda: ``We are not a perfect company''; paraphrased remarks on eagerness, model varieties and certification-stage burden, and on possibly overly stringent rules, while not condoning the violations.
\end{itemize}

\noindent\textbf{Sources.} C-0064: ``Results of manufacturers' investigation reports on irregularities in type designation applications'', MLIT (in Japanese); ``Results of Investigation Regarding Model Certification Applications'', Toyota Motor Corporation; ``Toyota apologizes for cheating on vehicle testing and halts production of three models'', Associated Press (via Spectrum News NY1). C-0065: the Toyota and AP items above. C-0066, C-0067: the AP item above; ``Toyota chairman Akio Toyoda apologises over the type designation application issue'', Car Watch (Impress, in Japanese). C-0068: the AP item above; ``Toyota Car Safety Certification Scandal Prompts Calls for Rules Review'', Bloomberg (via Claims Journal). C-0069: ``Investigation Report on Applications for Type Designation Submitted to the Ministry of Land, Infrastructure, Transport and Tourism'', Mazda Motor Corporation; the AP item above. C-0070: ``Discovery of inappropriate cases in four-wheel-vehicle type designation applications'', Honda Motor Co., Ltd.\ (in Japanese); the MLIT results release and AP item above. C-0071, C-0091: the MLIT results release above. C-0072: ``Toyota publishes investigation results on type designation applications: detailed press-conference report'', Toyota Times (in Japanese); the Car Watch item above; ``Toyota Halts Sales of Three JDM Models Over Crash Test Irregularities'', Motor1. C-0092: the Toyota Times item and Toyota results release above. C-0081: as in B.7.

\smallskip\noindent\textbf{B.10 Takata, 2016--2017 (adjudicated; M1, Create/Generate; conduct from around 2000)}\par

\begin{itemize}\setlength{\itemsep}{1pt}
\item \textbf{From around 2000 through 2015.} Takata submitted false and fraudulent inflator test reports to automaker customers, concealing ruptures during testing; admitted in its US guilty plea.
\item \textbf{May 2016.} NHTSA Amended Consent Order expanded recalls to all non-desiccated PSAN inflators: an estimated 35--40 million inflators added to about 28.8 million already under recall, phased through 2019.
\item \textbf{13 January 2017.} DOJ announced Takata's agreement to plead guilty to wire fraud with USD 1 billion in criminal penalties; an unsealed indictment charged three executives, who are presumed innocent.
\item \textbf{27 February 2017.} Guilty plea and sentence in the Eastern District of Michigan: USD 25 million fine, USD 975 million restitution, three years' probation, compliance monitor.
\item \textbf{November 2017.} NHTSA: 19 manufacturers, about 46 million inflators, an estimated 34 million US vehicles; ``the largest and most complex vehicle recalls in U.S.\ history''.
\end{itemize}

\noindent\textbf{Sources.} C-0111, C-0112: ``Takata Corporation Agrees to Plead Guilty and Pay \$1 Billion in Criminal Penalties for Airbag Scheme'', US Department of Justice (USAO Eastern District of Michigan); also ``Takata Corporation Pleads Guilty, Sentenced to Pay \$1 Billion in Criminal Penalties for Airbag Scheme'', US Department of Justice (via Internet Archive snapshot). C-0113: ``The State of the Takata Recalls'', NHTSA (via Internet Archive snapshot). C-0123: the USAO Eastern District of Michigan release above. C-0124: ``Takata Recall Expansion: What Consumers Need to Know'', NHTSA (via Internet Archive snapshot).

\smallskip\noindent\textbf{B.11 Volkswagen, 2015--2017 (adjudicated; M5, Create/Generate; affected vehicles sold since 2008)}\par

\begin{itemize}\setlength{\itemsep}{1pt}
\item \textbf{18 September 2015.} EPA notice of violation of the Clean Air Act to Volkswagen AG, Audi AG and Volkswagen Group of America: software detecting official testing, on-road NOx up to 40 times the standard, roughly 499,000 US cars; CARB issued an in-use compliance letter.
\item \textbf{22 September 2015.} VW ad-hoc disclosure: some eleven million vehicles worldwide with Type EA 189 engines; EUR 6.5 billion provision.
\item \textbf{28 June 2016.} Civil settlements of up to USD 14.7 billion covering buybacks, consumer compensation, mitigation and zero-emission investment for the 2.0-litre vehicles.
\item \textbf{11 January 2017.} VW agreed to plead guilty to three felony counts, with USD 1.5 billion in further civil resolutions; the US criminal case describes approximately 590,000 US vehicles.
\item \textbf{10 March and 21 April 2017.} Guilty plea entered; sentencing in the Eastern District of Michigan to a USD 2.8 billion criminal penalty, probation and an independent compliance monitor.
\item \textbf{Boundary note.} The evidence chain was facially complete and internally consistent; the falsity lay in the artefact's engineered behaviour under test.
\end{itemize}

\noindent\textbf{Sources.} C-0114: ``EPA, California Notify Volkswagen of Clean Air Act Violations'', US Environmental Protection Agency. C-0115: ``Volkswagen AG Agrees to Plead Guilty and Pay \$4.3 Billion in Criminal and Civil Penalties; Six Volkswagen Executives and Employees are Indicted'' and ``Volkswagen AG Sentenced in Connection with Conspiracy to Cheat U.S.\ Emissions Tests'', US Department of Justice (via Internet Archive snapshots); ``Volkswagen Agrees to Plead Guilty, Pay \$4.3 Billion in Criminal and Civil Penalties'', US Environmental Protection Agency. C-0116: ``DGAP-Adhoc: VOLKSWAGEN AG has issued the following information'', Volkswagen AG (via GlobeNewswire); ``The Emissions Issue: Volkswagen Group Annual Report 2015'', Volkswagen AG (via Internet Archive snapshot). C-0121: ``Volkswagen to Spend up to \$14.7 Billion to Settle Allegations of Cheating Emissions Tests and Deceiving Customers on 2.0 Liter Diesel Vehicles'', US Department of Justice (via Internet Archive snapshot). C-0122: the EPA notice-of-violation release and DOJ sentencing release above.

\smallskip\noindent\textbf{B.12 GM India, 2013 (company-acknowledged; M2, Verify/Reconcile)}\par

\begin{itemize}\setlength{\itemsep}{1pt}
\item \textbf{July 2013.} GM India recalled 114,000 Chevrolet Taveras (BS3 and BS4 diesel variants, 2005--2013), halting Tavera production and sales; GM stated the issues were not safety-related.
\item \textbf{Count conflict.} The transport minister put the recall at 160,000; GM maintained 114,000.
\item \textbf{2013.} GM acknowledged that an internal investigation found employees keeping fine-tuned lower-emission engines aside for installation during inspection, and dismissed employees including Sam Winegarden, vice-president for global engine engineering.
\item \textbf{October 2013.} A government inter-ministerial committee was reported to have found testing rules flouted and to have recommended ending the ten-day prior notice for conformity-of-production testing; the report itself is not on the verified record.
\end{itemize}

\noindent\textbf{Sources.} C-0117: ``General Motors India Recalls 114,000 Cars'', IndustryWeek (AFP); ``GM India crashes into Tavera mess, faces Rs 11 cr fine for corporate fraud'', Business Today; ``Tavera recall results in shake-up at General Motors'', Autocar Professional. C-0118: the Business Today item above; ``General Motors may be charged with corporate fraud over Tavera recall issue'', CarWale; ``GM Fires Employees over Indian Recall'', IndustryWeek (AFP). C-0128: the IndustryWeek (AFP) dismissals item above; ``GM Recalls 114,000 Taveras and 50-plus Executives'', Knowledge at Wharton; ``GM Sacks Employees in US and India after Probe Into Chevy SUV Recall'', IBTimes UK.

\smallskip\noindent\textbf{B.13 Hyundai/Kia, 2012--2014 (adjudicated; M2, Create/Generate)}\par

\begin{itemize}\setlength{\itemsep}{1pt}
\item \textbf{2 November 2012.} EPA announced that Hyundai and Kia would lower fuel-economy estimates on most model year 2012--2013 vehicles after Ann Arbor laboratory testing diverged from submitted data; cars on dealer lots relabelled.
\item \textbf{3 November 2014.} Consent decree lodged in the US District Court for the District of Columbia: USD 100 million civil penalty, forfeiture of 4.75 million greenhouse-gas credits valued over USD 200 million, about USD 50 million in compliance measures.
\item \textbf{Scale.} Close to 1.2 million vehicles; approximately 4.75 million metric tons of greenhouse gases in excess of certification.
\end{itemize}

\noindent\textbf{Sources.} C-0119: ``Hyundai/Kia to Correct Overstated MPG Claims as Result of EPA Investigation'', US EPA. C-0120, C-0129: ``United States Reaches Settlement with Hyundai and Kia in Historic Greenhouse Gas Enforcement Case'', US EPA.

\section{AIAG PPAP and VDA Volume 2 PPA, extended comparison}\label{app:cmp}
This appendix extends the comparison of Table 2 with history and an analytic reading of the two regimes' design choices.

History. VDA Volume 2 PPA (the PPF procedure in German usage), ``Quality Assurance of Supplies'', has defined the German industry's production process and product approval since its first publication in 1975. The current edition is the 7th revised edition of January 2026, and the procedure is the functional counterpart of AIAG PPAP. The evidence instrument German OEMs traditionally request is the initial sample inspection report (ISIR; German Erstmusterpr\"ufbericht, EMPB); current editions frame the evidence as PPA deliverables and PPA samples.

Submission depth: levels versus agreement. AIAG fixes five submission levels, set at the customer's discretion, with Level 3 (warrant, samples, complete supporting data) the common default. VDA eliminated submission levels in the 6th edition (2020); the scope, content, and schedule of the submission are instead agreed between organisation and customer in the PPA agreement. The design difference is real: AIAG standardises disclosure depth into a five-value enumeration, while VDA moves the same decision into a negotiated bilateral document. For an information-system view of the chain, the AIAG choice is machine-legible per submission, whereas the VDA choice relocates the configuration into contract-like records that vary pair by pair.

Approval states: dispositions versus release. AIAG defines three dispositions: approved; interim approval, permitting shipment for production requirements on a limited time or piece-quantity basis while full approval is pending; and rejected, requiring correction and resubmission. Interim approval is customer-conditioned in practice: Ford requires an approved and authorised WERS alert for interim PSW approval, the alert itself being the authorisation to ship parts. VDA eliminated conditional approvals in the same 6th edition. The asymmetry is analytically useful. AIAG's interim state makes shipping-before-full-approval a first-class, recorded evidence state; the VDA procedure since 2020 declines to encode partial approval at all, leaving any such arrangement outside the standard state model.

Risk basis. The 7th edition of VDA Volume 2 adopts a risk-based approach in which the scope of the PPA procedure is determined by the risk level of the products to be released. AIAG's level mechanism also modulates evidential burden, but by customer discretion per part rather than by a declared risk basis.

Convergence. Methodological convergence is already delivered in one artefact family: AIAG and VDA jointly published the harmonised AIAG and VDA FMEA Handbook (1st edition, June 2019), replacing the previously separate FMEA methodologies with a common, more structured seven-step approach. The 7th edition's foreword records that harmonisation with AIAG PPAP was discussed, the German industry being open to a common international standard in the long term. A single production-approval regime is therefore a stated long-term prospect on the German side, not this paper's speculation.

\smallskip\noindent Table C1. Extended comparison of the two production-approval regimes.

{\footnotesize\centering
\begin{longtable}{p{2.4cm}p{6.3cm}p{6.3cm}}
\toprule
Dimension & AIAG PPAP & VDA Volume 2 PPA \\
\midrule\endhead
Lineage & AIAG PPAP manual, operating under IATF 16949 together with CSRs & ``Quality Assurance of Supplies'', first published 1975; 7th revised edition, January 2026 \\
Evidence instrument & 18-element package summarised and gated by the PSW & PPA deliverables and PPA samples; traditionally the ISIR/EMPB \\
Submission depth & Levels 1 to 5 at customer discretion; Level 3 the common default & Levels eliminated in the 6th edition (2020); scope, content, and schedule per the PPA agreement \\
Approval states & Approved, interim approval, rejected; interim conditioned by CSR (Ford's WERS alert) & Conditional approvals eliminated in the 6th edition (2020) \\
Scoping logic & Fixed element set with level-based disclosure & Risk-based: procedure scope set by product risk level (7th edition) \\
Convergence signal & Joint AIAG and VDA FMEA Handbook, June 2019 & 7th edition foreword: PPAP harmonisation discussed; openness to a common international standard long term \\
\bottomrule
\end{longtable}
}

\section{Nearest-neighbour literature table}\label{app:nn}
This appendix cashes the survey-scoped emptiness claim of Section 2. Of the 53 works surveyed, the eight below come closest to this paper's intersection; the final column states the specific delta between each work and this paper.

{\footnotesize\centering
\begin{longtable}{p{2.4cm}p{1.0cm}p{2.5cm}p{4.6cm}p{4.6cm}}
\toprule
Work & Year & Venue & What it does & Delta to this paper \\
\midrule\endhead
Chi, Sigmund and Astardi \cite{Chi2020} & 2020 & Reliability Engineering and System Safety & Classifies root causes of 345 NHTSA passenger-vehicle recalls into defect classes, translated to FMEA & Closest taxonomy: the classified objects are product defects behind recalls, not compliance-documentation failures; no evidence-chain systematisation \\
Haeckel, Husung and Wuensche \cite{Haeckel2025} & 2025 & International Journal of Conformity Assessment & Identifies technologies to automate one conformity-of-production verification step at BMW: partially manual, randomly sampled checks of component identification numbers and homologation labels & Closest engineering of the chain: a single verification step, with the surrounding chain taken as given; no failure record, no taxonomy \\
Se\v{n}ov\'{a}, \v{S}tofov\'{a}, Szarysz\'{o}v\'{a} and Dugas \cite{Senova2021} & 2021 & Management Systems in Production Engineering & ISO/TS 16949 process approach with PPAP as the approval mechanism for parts in series production & Closest quality-management work: PPAP as a compliance step inside a certification frame; no information-system architecture, no failure taxonomy \\
Hashmi, Governatori, Lam and Wynn \cite{Hashmi2018} & 2018 & Knowledge and Information Systems & Authoritative survey of business-process compliance: design-time, run-time and auditing approaches to checking processes against regulatory obligations & Closest information-systems machinery: domain-generic, with no automotive instantiation \\
Hara and Fujimura \cite{HaraFujimura2024} & 2024 & Technology in Society & Matched-pair case study attributing quality-data falsification at Japanese firms to inadequate capability, organisational myopia and standards uncertainty & Closest failure analysis: explains why fraud occurs at firm level; does not systematise the chain the fraud traversed or taxonomise its mechanisms \\
Rozinat and van der Aalst \cite{RozinatAalst2008} & 2008 & Information Systems & Conformance checking of process logs against process models: fitness and appropriateness & Sharpest boundary marker: detects control-flow deviation; a fabricated value inside a procedurally complete record is conformant by construction \\
MacNeil \cite{MacNeil2002} & 2002 & Archivaria & InterPARES Authenticity Task Force requirements for electronic-record authenticity & Conceptual mirror image: states what trustworthy records need where this paper's taxonomy states how they break; no domain instantiation \\
Patro, Ahmad, Yaqoob, Salah and Jayaraman \cite{Patro2021} & 2021 & IEEE Access & Ethereum/IPFS system for recall visibility and fraud prevention in the automotive supply chain & Closest automotive document infrastructure: solution-first, post-market recall logistics; no mapping of the deployed pre-market evidence chain, no failure record \\
\bottomrule
\end{longtable}
}

No surveyed work satisfies both prongs of the gap: systematising the automotive supplier-quality evidence chain as an information system and offering a failure taxonomy over its documented failures.

\end{document}